\documentclass[pdflatex,sn-mathphys-num]{sn-jnl}

\usepackage[T1]{fontenc}
\usepackage[utf8]{inputenc}
\usepackage{lineno,hyperref}
\modulolinenumbers[20]
\usepackage[linesnumbered,ruled,vlined]{algorithm2e}
\usepackage{graphicx, subfigure}
\usepackage{color}
\usepackage{tcolorbox}
\usepackage{booktabs}
\usepackage{longtable}
\usepackage{amssymb}
\usepackage{amsmath}
\usepackage{dsfont}
\usepackage{bbm}
\usepackage{lipsum}
\usepackage{enumitem}
\usepackage{multirow}
\usepackage{listings}
\usepackage[table]{xcolor}

\usepackage{listings}
\usepackage{xcolor}

\lstdefinelanguage{nasm}{
  morekeywords={mov,xor,add,sub,inc,dec,cmp,jmp,jl,jle,jg,jge,je,jne,jz,jnz,
                push,pop,call,ret,nop,lea,test,and,or,shl,shr,imul,idiv,
                short,near,far,byte,word,dword,qword,ptr},
  morekeywords=[2]{eax,ebx,ecx,edx,esi,edi,ebp,esp,
                   rax,rbx,rcx,rdx,rsi,rdi,rbp,rsp,
                   ax,bx,cx,dx,al,bl,cl,dl,ah,bh,ch,dh},
  sensitive=false,
  morecomment=[l]{;},
  morestring=[b]",
  morestring=[b]',
  alsodigit={0x}
}

\lstdefinestyle{nasmstyle}{
  language=nasm,
  basicstyle=\ttfamily\scriptsize,
  keywordstyle=\color{blue!70!black}\bfseries,
  keywordstyle=[2]\color{purple!80!black},
  commentstyle=\color{gray}\itshape,
  stringstyle=\color{orange!80!black},
  numberstyle=\tiny\color{gray},
  frame=single,
  rulecolor=\color{black!50},
  showstringspaces=false,
  breaklines=true,
  columns=fullflexible,
  keepspaces=true,
  tabsize=2,
  xleftmargin=2pt,
  xrightmargin=2pt
}

\usepackage{tikz}
\usepackage{adjustbox}
\usepackage{amssymb}
\usetikzlibrary{shapes.geometric, arrows.meta, positioning, fit, backgrounds, shadows, calc, patterns, decorations.pathmorphing}

\definecolor{techblue}{RGB}{230, 240, 250}       
\definecolor{techborder}{RGB}{70, 130, 180}      
\definecolor{processyellow}{RGB}{255, 242, 204}  
\definecolor{securegreen}{RGB}{200, 230, 200}    
\definecolor{alertred}{RGB}{245, 200, 200}       
\definecolor{darkslate}{RGB}{50, 50, 60}         
\definecolor{softgray}{RGB}{245, 245, 245}       

\usepackage[numbers]{natbib}

\graphicspath{{figs/}}

\begin{document}

\title[AsmRAG]{AsmRAG: LLM-Driven Malware Detection by Retrieving Functionally Similar Assembly Code} 

\author*[1]{\fnm{ElMouatez Billah} \sur{Karbab}}\email{mouatez@karbab.net}
\affil*[1]{\orgname{Joaan Bin Jassim Academy for Defence Studies}, \orgaddress{\city{Al-Khor}, \country{Qatar}}}



\abstract{
Deep learning malware detectors achieve high classification accuracy but suffer from severe interpretability limitations, typically returning probabilistic verdicts that lack forensic context. We introduce \texttt{AsmRAG}, a framework performing malware analysis through \textit{Assembly-Level Retrieval-Augmented Generation}. Unlike classifiers built on global statistical features, \texttt{AsmRAG} reformulates detection as an evidence-based retrieval task. The system uses a code-specialized Large Language Model (LLM) to analyze assembly functions and convert them into semantic embeddings. This process constructs a searchable knowledge base resilient to syntactic obfuscation. For inference, we propose a \textit{Density-Weighted Anchor Selection} mechanism that isolates the primary unit of malicious logic within a binary to extract verifiable forensic evidence and resist evasion attempts. Testing on a curated dataset of 40k binaries shows \texttt{AsmRAG} reaching a \textit{detection F1-score of 96\%} alongside a family \textit{attribution F1-score of 95\%}. Comparisons confirm this semantic retrieval approach remains robust against metamorphic obfuscation. When holistic baselines (EMBER and ResNeXt) degrade, our methodology gives Security Operations Centers a transparent and reliable alternative.
}


\maketitle

\section{Introduction}
\label{sec:introduction}
Deep learning provides stronger defenses against automated polymorphism than static signatures~\cite{Saxe2015Deep, saqib2023comprehensive}, yet it introduces significant limitations in interpretability and robustness. Modern ``black box'' models (e.g., EMBER~\cite{anderson2018ember}, ResNeXt~\cite{He2024ResNeXtAM}) give Security Operations Center analysts opaque probabilistic verdicts. These systems also remain vulnerable to global obfuscation techniques, such as dead-code insertion, which alter file-level statistics without changing local semantics. We propose shifting malware analysis from closed-set \textit{classification} to open-set \textit{evidence-based retrieval}. This approach prioritizes identifying functionally equivalent malicious logic rather than relying on statistical approximation.

We introduce \texttt{AsmRAG}, a framework that applies \textit{Retrieval-Augmented Generation (RAG)} to low-level assembly code. Unlike feature-based classifiers, \texttt{AsmRAG} moves beyond binary labels to retrieve specific, verifiable code snippets from a knowledge base of known threats. To bridge the ``Semantic Gap'' where different assembly instructions execute identical logic, we use a code-specialized Large Language Model (LLM) to generate semantic embeddings. These embeddings map functionally equivalent functions to close coordinates in a vector space, which keeps the retrieval process resilient against syntactic obfuscation. This method establishes direct \textit{Forensic Explainability}. The system isolates the most representative ``Anchor Function'' inside a suspect binary and compares it against a retrieved ``proof'' sample. Through this mechanism, \texttt{AsmRAG} delivers clear, interpretable evidence for malware attribution.

This paper makes the following contributions:
\begin{enumerate}
  \item \textbf{Assembly RAG Detection Architecture.} We design \texttt{AsmRAG}, a framework that uses a RAG architecture for malware detection at the \textit{function level}. Shifting the analysis from global file statistics to targeted semantic retrieval grounds the detection process in atomic forensic evidence.

  \item \textbf{Noise-Resilient Embedding Strategy.} We present a method to generate dense vector representations by extracting assembly functions, minimizing canonicalization steps, and embedding the results with a pre-trained LLM. We pair this with a new \textbf{Semantic Library Filtering} mechanism to discard the benign boilerplate code that typically overwhelms function-level classifiers.

  \item \textbf{Density-Weighted Anchor Attribution.} We engineer a scoring system to identify the ``Anchor Function'' ($\mathbf{q}^*$) using the density of its malicious neighborhood. This isolation drives the generation of targeted natural language explanations, translating high-dimensional vector similarity into human-readable insights.

  \item \textbf{Evaluation and Comparison.} We test our approach on a curated dataset of 40,814 binaries spanning 17 malware families. \texttt{AsmRAG} reaches a detection F1-score of \textbf{96.5\%} alongside an attribution F1-score of \textbf{0.95}. We show improved robustness against adversarial compilation compared to industry baselines (EMBER~\cite{anderson2018ember}, ResNeXt~\cite{He2024ResNeXtAM}). These results confirm the stability of semantic retrieval against common evasion attempts.
\end{enumerate}

The remainder of this paper is organized as follows: Section~\ref{sec:background} defines the threat model and the semantic gap. Section~\ref{sec:system_design} details the system architecture and describes the embedding pipeline. Section~\ref{sec:evaluation} presents empirical results on detection performance and robustness. We discuss system limitations in Section~\ref{sec:discussion} and conclude in Section~\ref{sec:conclusion}.

\section{Background and Motivation}
\label{sec:background}

Learning-based malware detection systems achieve high \textit{discriminative accuracy} but often suffer from severe deficits in \textit{forensic interpretability}. In this section, we formalize the adversarial landscape and argue that resilient detection requires a shift from probabilistic classification to semantic retrieval.

\subsection{The Adversarial Landscape: Semantic Invariance}
\label{subsec:threat_model}

We model the defense of binary executables as a game defined over the space of programs $\mathcal{P}$. The adversary seeks to generate a variant $P'$ from a malicious seed $P_{mal}$ that evades detection while preserving the original malicious payload.

We define this formally through \textit{Trace Equivalence}. Let $\Sigma$ be the set of observable architectural states (e.g., memory writes, syscalls). A given program $P$ induces a set of execution traces $\text{Tr}(P) \subseteq \Sigma^*$. An adversarial transformation function $\mathcal{T}: \mathcal{P} \to \mathcal{P}$ is valid if and only if it preserves observational equivalence:
\begin{equation}
  \forall \tau \in \text{Tr}(P), \exists \tau' \in \text{Tr}(\mathcal{T}(P)) : \tau \approx \tau'
\end{equation}
Here, $\approx$ denotes semantic equivalence, meaning the sequence of effective state changes remains constant despite syntactic perturbations.

\vspace{1mm}
\noindent \textbf{The Syntactic Noise Problem.} 
To achieve evasion, adversaries exploit the inherent redundancy of the Instruction Set Architecture (ISA). They inject \textit{syntactic noise} by inserting dead code, substituting instructions, and flattening control flow. This manipulation expands the high-dimensional volume of the malware family, pushing $P'$ away from $P$ in the raw feature space (e.g., byte n-grams or opcode histograms) without altering the underlying logic graph.

\subsubsection{The Boundary Sensitivity Problem}
Discriminative models rely on statistical approximations, making them inherently sensitive to the geometry of their decision boundaries. Obfuscation techniques act as adversarial perturbations designed to push a sample $\mathbf{x}_{P'}$ across these boundaries. Because standard models operate on ``surface'' features such as token frequency or byte entropy, the distance in feature space $\Delta(\mathbf{x}_P, \mathbf{x}_{P'})$ can grow arbitrarily large despite semantic identity. Consider the example in Listing~\ref{lst:motivational}. A sequence-based model (e.g., LSTM~\cite{Hochreiter1997}) trained on $P$ will likely assign a low probability to $P'$ because of the added noise (\texttt{nop}) and the register swap ($\texttt{eax} \to \texttt{ebx}$). The model effectively treats the obfuscated variant as a distinct, unknown entity.

\begin{figure}[h]
\begin{lstlisting}[style=nasmstyle]
; (a) Reference Logic (Seed)
loc_start:
mov  eax, [ebp+var_4]   ; Load accumulator
xor  eax, 0x5A          ; Payload: Encrypt
mov  [ebp+var_4], eax   ; Store
inc  ecx                ; Update iterator
cmp  ecx, 100           ; Loop condition
jl   short loc_start
\end{lstlisting}

\vspace{0.5em}

\begin{lstlisting}[style=nasmstyle]
; (b) Obfuscated Variant P'
loc_new:
nop                     ; [Noise] Dead code
mov  ebx, [ebp+var_4]   ; [Reg Swap] eax -> ebx
xor  ebx, 90            ; [Synonym] 0x5A == 90
mov  [ebp+var_4], ebx   ; Store (same slot)
add  ecx, 1             ; [Subst] inc -> add
cmp  ecx, 0x64          ; [Synonym] 100 -> 0x64
jl   short loc_new      ; [Rename] label only
\end{lstlisting}
\caption{Semantic Invariance amidst Syntactic Variance. The adversary alters
the syntax (registers, opcodes, literals) and topology (dead code insertion,
label renaming) while preserving the data-flow graph over the live
variables.}
\label{lst:motivational}
\end{figure}


\subsection{The Interpretability Gap in Deep Learning}
\label{subsec:interpretability_gap}

Adopting Deep Learning (DL) models~\cite{He2024ResNeXtAM,karbab2023swiftr} builds necessary resilience against such obfuscation. This shift, however, introduces a critical operational flaw commonly known as the ``Black Box'' problem.

Let $f_\theta: \mathcal{P} \to [0,1]$ be a neural detector parameterized by $\theta$. Within a Security Operations Center (SOC), a simple prediction like $f_\theta(P') > 0.99$ is insufficient for practical response. Analysts require \textit{attribution}: they must know exactly \textit{why} a binary is malicious rather than just accepting \textit{that} it is malicious.

\subsubsection{Failure of Attribution}
Standard discriminative models define a decision boundary to partition the available feature space. They compress input data into a latent vector $z$, effectively destroying any structural mapping back to the original assembly code. When $f_\theta$ successfully detects an obfuscated variant $P'$, the system remains unable to localize the specific granular functions responsible for that malicious verdict. This structural disconnect creates a formal \textit{Interpretability Gap}:
\begin{equation}
  \text{Gap}(P') = \mathcal{H}(Y | P') - \mathcal{H}(Y | P', \mathcal{E})
\end{equation}
Here, $\mathcal{H}$ represents entropy, $Y$ denotes the ground truth label, and $\mathcal{E}$ indicates the forensic evidence (such as a reference to a known non-obfuscated function). Current state-of-the-art models~\cite{He2024ResNeXtAM,karbab2023swiftr} successfully minimize prediction error yet consistently fail to provide this critical evidence $\mathcal{E}$.

\begin{figure*}[t]
\centering
\includegraphics[width=12cm]{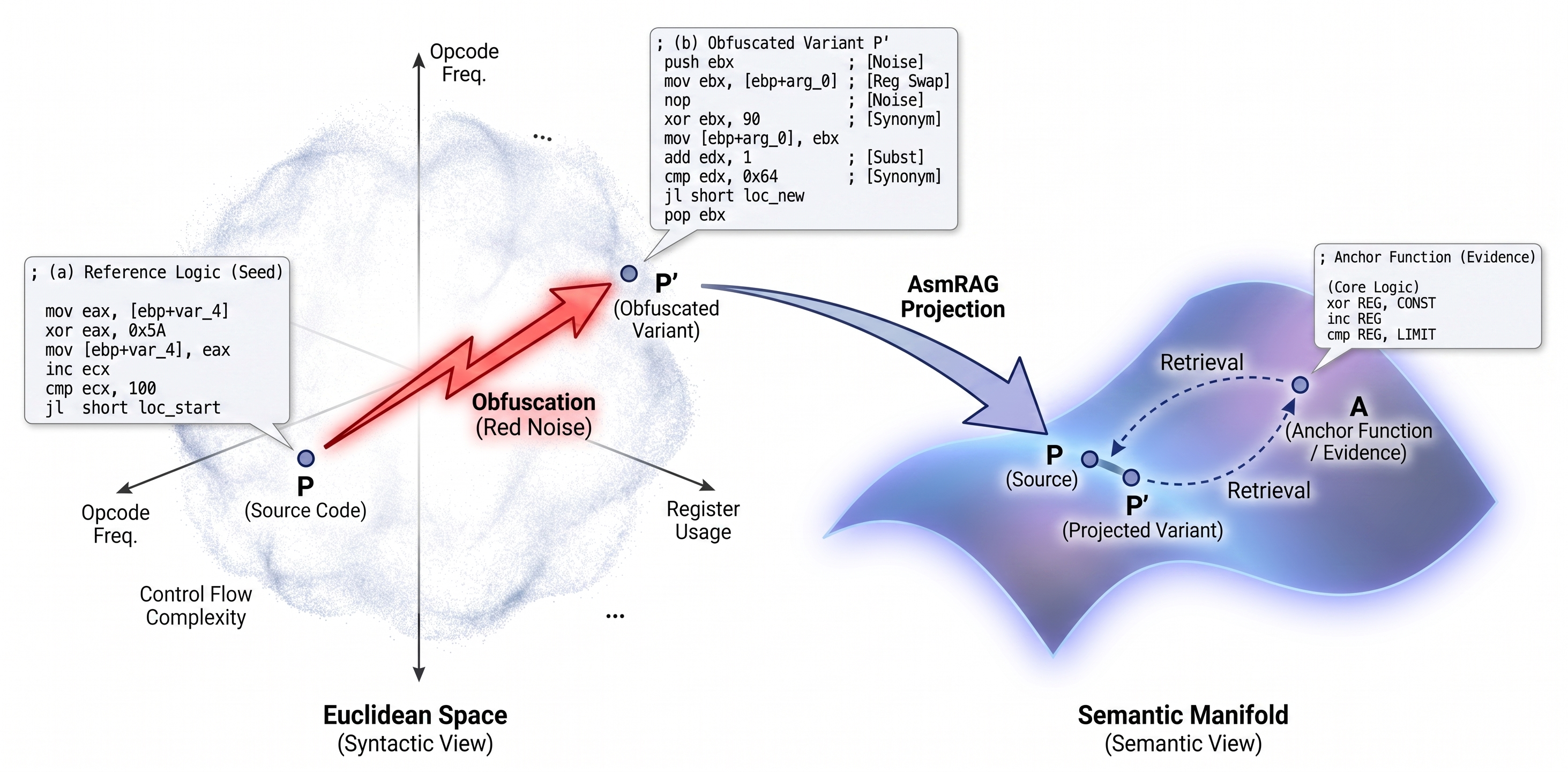} 
\caption{\textbf{The Manifold Hypothesis for Forensics.} While obfuscation (red noise) pushes the variant $P'$ far from the source $P$ in Euclidean space, they remain proximal on the Semantic Manifold. \texttt{AsmRAG} projects inputs to this manifold to retrieve the ``Anchor Function'' (evidence).}
\label{fig:manifold}
\end{figure*}

\subsection{Retrieval as Forensic Grounding}
\label{subsec:retrieval_grounding}

We address this gap by reframing detection as a \textbf{Nearest Neighbor Retrieval} problem on a semantic manifold. 

\subsubsection{The Manifold Hypothesis}
We posit that valid assembly functions reside on a low-dimensional manifold $\mathcal{M} \subset \mathbb{R}^d$ embedded within the high-dimensional space of possible byte sequences. Obfuscation thus acts as a perturbation vector $\epsilon$ that displaces $P$ from the ``canonical'' manifold. 

Traditional classifiers attempt to delineate complex boundaries around these noisy regions. Instead, \texttt{AsmRAG} (see Figure \ref{fig:manifold}) learns a projection $\Phi$ that maps both the pristine malware $P$ and the obfuscated variant $P'$ to the same resilient neighborhood on $\mathcal{M}$:
\begin{equation}
  ||\Phi(P) - \Phi(P')||_2 < \delta \ll ||\Phi(P) - \Phi(\text{Benign})||_2
\end{equation}

\subsubsection{Anchor Function Attribution}
By utilizing a Retrieval-Augmented Generation (RAG) architecture, we do not merely classify $P'$; we retrieve $P$. This process identifies the \textbf{``Anchor Function''}, the specific code segment in the retrieved binary that shares the highest semantic similarity with the suspicious input.

This mechanism bridges the Interpretability Gap. Instead of an opaque probability score, the system provides a \textit{differential explanation}: ``This binary is malicious because Function A contains logic semantically isomorphic to the encryption routine in Family X, despite the insertion of superfluous instructions.'' This shifts the paradigm from \textit{probabilistic inference} to \textit{evidence-based retrieval}.

\subsubsection{From Cognitive Abstraction to Vector Projection}
\label{subsub:human_intuition}

Conventional automated detection enables rapid processing but suffers from severe deficits in robustness compared to human analysis. This divergence stems from the level of abstraction applied. A human reverse engineer analyzing Snippet (b) performs \textit{cognitive abstraction} through the following steps:
\begin{enumerate}
  \item \textbf{Noise Filtering:} The analyst mentally discards the \texttt{nop} instruction as functionally irrelevant.
  \item \textbf{Normalization:} They recognize that \texttt{inc ecx} and \texttt{add edx, 1} represent semantically isomorphic operations for incrementing a loop counter.
  \item \textbf{Data Flow Tracking:} They track the data flow into register \texttt{ebx}, identifying it as a temporary storage container equivalent to \texttt{eax} in Snippet (a), independent of the register's nomenclature.
\end{enumerate}

Through this process, the analyst concludes that the \textit{semantics} (the functional intent of XORing a buffer in a loop) remain invariant, identifying the code as a variant of the known malware family. This manual cognitive process, however, is slow, resource-intensive, and unscalable against the sheer volume of modern threats.

\texttt{AsmRAG} operationalizes this expert intuition at scale. The system extracts and analyzes assembly with a Large Language Model pre-trained on vast code repositories, allowing it to learn an automated abstraction function. The model projects both the concise reference (a) and the noisy variant (b) to proximal points in a high-dimensional vector space ($\mathbf{v}_a \approx \mathbf{v}_b$). This geometry enables the system to resist syntactic obfuscation in a manner analogous to human cognition. It retrieves the original malicious code as verifiable forensic evidence without incurring the prohibitive costs of manual inspection or heavy-weight symbolic execution.

\section{AsmRAG System Design}
\label{sec:system_design}

Current detection models achieve high accuracy but suffer from severe limitations in human interpretability. We introduce \texttt{AsmRAG}, a resilient paradigm designed to bridge this gap. We structure this granular system by constructing the Knowledge Base, performing Evidence-Based Analysis, and executing Generative Reporting.

\begin{figure*}[t]
\centering
\resizebox{1.1\textwidth}{!}{
\begin{tikzpicture}[
    node distance=1.2cm and 1.0cm, 
    font=\sffamily\footnotesize,
    >=Stealth,
    block/.style={
        rectangle, 
        draw=darkslate, 
        fill=white, 
        rounded corners=3pt, 
        align=center, 
        text width=2.5cm, 
        minimum height=1cm, 
        drop shadow, 
        thick
    },
    process/.style={
        rectangle, 
        draw=techborder, 
        fill=techblue, 
        rounded corners=3pt, 
        align=center, 
        text width=2.2cm, 
        minimum height=1.2cm, 
        drop shadow
    },
    model/.style={
        rectangle, 
        draw=orange!80!black, 
        fill=processyellow, 
        rounded corners=3pt, 
        align=center, 
        text width=2.2cm, 
        minimum height=1.2cm, 
        drop shadow
    },
    db/.style={
        cylinder, 
        shape border rotate=90, 
        draw=darkslate, 
        fill=white, 
        aspect=0.25, 
        minimum height=1.5cm, 
        minimum width=1.5cm, 
        align=center, 
        drop shadow
    },
    decision/.style={
        diamond, 
        draw=red!80!black, 
        fill=alertred!30, 
        align=center, 
        aspect=2.5, 
        inner sep=0pt, 
        text width=2cm, 
        drop shadow
    },
    labelnode/.style={
        text=darkslate, 
        font=\bfseries\small, 
        align=left, 
        anchor=west
    },
    line/.style={draw, ->, thick, darkslate},
    dashedline/.style={draw, ->, dashed, thick, darkslate}
]


\node[labelnode] (stage1_label) at (0, 0) {Stage 1: Knowledge Base Construction};

\node[block, below right=0.2cm and 0cm of stage1_label.south west, anchor=north west, fill=none, draw=none, text width=2.5cm] (dataset) {\textbf{Raw Dataset}\\\small(Malware + Benign)};

\node[process, right=0.6cm of dataset] (disasm) {Disassembler\\(IDA/Ghidra)};
\node[process, right=0.6cm of disasm] (filter) {Library Filter\\(Blocklist $\mathcal{L}$)};
\node[process, right=0.6cm of filter] (canon) {Minimal\\Canonicalization};

\node[model, right=0.6cm of canon] (qwen) {\textbf{Qwen3-Embedding}\\(Zero-Shot Emb)};

\node[db, right=0.8cm of qwen] (faiss) {\textbf{FAISS Index}\\($\mathcal{D}_{KB}$)};


\node[labelnode, below=2.5cm of stage1_label] (stage2_label) {Stage 2: Evidence-Based Analysis};

\node[block, below right=0.2cm and 0cm of stage2_label.south west, anchor=north west, fill=none, draw=none, text width=2.5cm] (unknown) {\textbf{Unknown Sample}\\\small $S$};

\node[process, below=2.0cm of disasm] (disasm_on) {Disassembler};
\node[process, below=2.0cm of filter] (filter_on) {Library Filter};
\node[process, below=2.0cm of canon] (canon_on) {Minimal\\Canonicalization};
\node[model, below=2.0cm of qwen] (qwen_on) {\textbf{Qwen3-Embedding}\\(Inference)};

\node[process, below=1.8cm of faiss, fill=securegreen!20, draw=green!40!black] (knn) {Top-K Retrieval\\(Cosine Sim)};

\node[process, right=0.8cm of knn, text width=2.2cm] (voting) {Distance-Weighted\\Voting ($\alpha_i$)};

\node[decision, below=1.0cm of voting] (threshold) {$\Omega(S) > \tau_{file}$?};

\node[block, left=1.2cm of threshold, fill=alertred, draw=red!80!black] (malicious) {Malicious\\\& Family $C$};
\node[block, right=1.2cm of threshold, fill=securegreen, draw=green!40!black] (benign) {Benign};


\draw[line] (dataset) -- (disasm);
\draw[line] (disasm) -- (filter);
\draw[line] (filter) -- (canon);
\draw[line] (canon) -- (qwen);
\draw[line] (qwen) -- node[above, font=\scriptsize] {$\mathbf{v}_f$} (faiss);

\draw[line] (unknown) -- (disasm_on);
\draw[line] (disasm_on) -- (filter_on);
\draw[line] (filter_on) -- (canon_on);
\draw[line] (canon_on) -- (qwen_on);
\draw[line] (qwen_on) -- node[above, font=\scriptsize] {$Q_S$} (knn);

\draw[line] (faiss) -- node[right, font=\scriptsize] {ANN Search} (knn);

\draw[line] (knn) -- node[above, font=\scriptsize] {$\mathcal{N}_k$} (voting);
\draw[line] (voting) -- (threshold);

\draw[line] (threshold.west) -- node[above, font=\scriptsize] {Yes} (malicious.east);
\draw[line] (threshold.east) -- node[above, font=\scriptsize] {No} (benign.west);

\begin{scope}[on background layer]
    \node[fit=(stage1_label)(dataset)(faiss)(disasm), draw=darkslate!30, dashed, fill=softgray, rounded corners, inner sep=10pt] {};
    
    \node[fit=(stage2_label)(unknown)(benign)(malicious)(knn)(voting), draw=darkslate!30, dashed, fill=white, rounded corners, inner sep=10pt] {};
\end{scope}

\end{tikzpicture}
}
\caption{The \texttt{AsmRAG} System Architecture. The pipeline minimizes preprocessing, relying on the Qwen3-Embedding model to embed raw assembly semantics. The online phase retrieves evidence from the FAISS knowledge base to perform weighted detection.}
\label{fig:main_architecture}
\end{figure*}
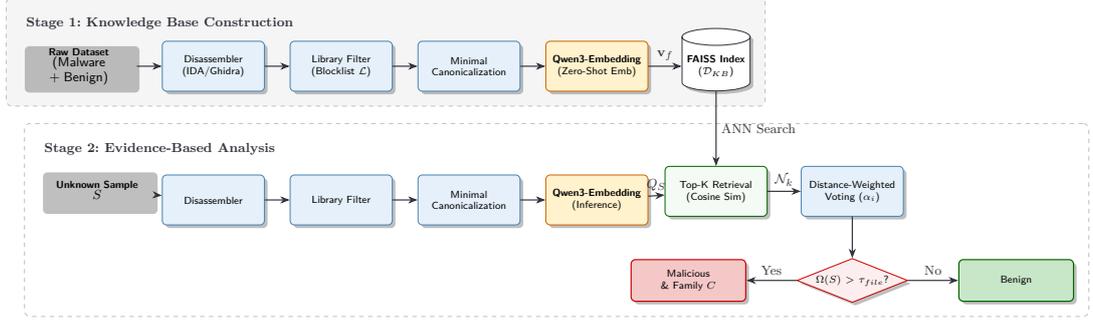

\begin{figure}[t]
\centering
\resizebox{0.6\textwidth}{!}{
\begin{tikzpicture}[
    node distance=1cm and 1cm,
    font=\sffamily\footnotesize,
    >=Stealth,
    data/.style={
        circle, 
        draw=techborder, 
        fill=techblue, 
        minimum size=0.5cm, 
        inner sep=0pt
    },
    anchornode/.style={
        circle, 
        draw=red!80!black, 
        fill=alertred, 
        text=darkslate, 
        font=\bfseries,
        minimum size=0.6cm, 
        inner sep=0pt, 
        drop shadow
    },
    process/.style={
        rectangle, 
        draw=darkslate, 
        rounded corners, 
        fill=white, 
        drop shadow, 
        minimum height=0.8cm, 
        text width=3cm, 
        align=center
    },
    llm/.style={
        rectangle, 
        draw=orange!80!black, 
        top color=processyellow, 
        bottom color=processyellow!80!orange, 
        rounded corners, 
        drop shadow, 
        minimum height=1cm, 
        minimum width=3cm, 
        text width=3cm, 
        align=center
    },
    human/.style={
        circle, 
        draw=darkslate, 
        fill=softgray, 
        minimum size=1.2cm, 
        align=center, 
        drop shadow,
        font=\scriptsize
    },
    line/.style={draw, ->, thick, darkslate},
    loopline/.style={draw, ->, thick, securegreen!80!black, dashed}
]

\node[process] (sample) {Flagged Sample $S$\\Functions:};

\node[data, below left=0.3cm and 0.2cm of sample.south] (f1) {};
\node[data, right=0.2cm of f1] (f2) {};
\node[anchornode, right=0.2cm of f2] (fstar) {$\mathbf{q}^*$};
\node[right=0.1cm of fstar, font=\scriptsize, text=alertred!80!black] {Anchor};

\node[process, below=1.5cm of sample, text width=4cm] (context) {Targeted Retrieval\\(Family $C_{best}$ Subspace)};
\draw[line] (fstar) -- (context);

\node[process, below left=0.8cm and -0.5cm of context, fill=alertred!20, text width=2cm] (q_node) {Query $\mathbf{q}^*$};
\node[process, below right=0.8cm and -0.5cm of context, fill=techblue, text width=2cm] (v_node) {Evidence $\mathbf{v}_{proof}$};

\draw[line] (context.south) -- (q_node.north);
\draw[line] (context.south) -- (v_node.north);

\node[llm, below=2.0cm of context] (generator) {\textbf{LLM Generator}\\(Explain Functional Logic)};

\draw[line] (q_node.south) -- (generator.north west);
\draw[line] (v_node.south) -- (generator.north east);

\node[process, below=0.8cm of generator] (explanation) {Explanation $\mathcal{E}$};
\draw[line] (generator) -- (explanation);

\node[human, right=1.2cm of explanation] (analyst) {Analyst\\/ Auto};
\draw[line] (explanation) -- (analyst);

\node[process, right=0.6cm of v_node, fill=securegreen, draw=securegreen!50!black] (kb_update) {Update KB\\($\mathcal{D}_{KB} \cup \mathbf{q}^*$)};

\draw[line] (analyst) -| node[pos=0.2, above, font=\scriptsize] {Confirm} (kb_update);

\draw[loopline] (kb_update.north) |- ([yshift=0.5cm]context.north) -| (context.north);

\node[font=\scriptsize\bfseries, text=securegreen!80!black, above right=0.5cm and 0.5cm of context.north east] {Active Learning};

\end{tikzpicture}
}
\caption{The Active Explanation Loop. \texttt{AsmRAG} selects the single most critical ``Anchor Function'' $\mathbf{q}^*$, generates a specific explanation against proof $\mathbf{v}_{proof}$, and ingests confirmed detections back into the database to enhance future resilience.}
\label{fig:active_learning}
\end{figure}
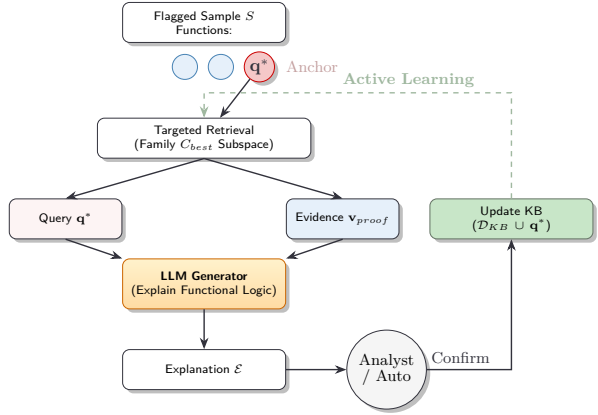

\subsection{Stage 1: Knowledge Base Construction}
The foundation of our system is a searchable index of semantic assembly embeddings, curated to maximize signal integrity and minimize noise.

\begin{algorithm}[h!]
\SetAlgoLined
\DontPrintSemicolon
\footnotesize
\caption{AsmRAG Inference and Anchor Selection}
\label{alg:asmrag_inference}

\KwIn{Unknown Binary $S$, Knowledge Base $\mathcal{D}_{KB}$, Blocklist $\mathcal{L}$}
\KwIn{Thresholds $\tau_{func}, \tau_{file}$}
\KwOut{Verdict, Family $C_{best}$, Explanation $\mathcal{E}$}

\tcp{Stage 1: Preprocessing \& Embedding}
$F_{raw} \leftarrow \text{Disassemble}(S)$\;
$Q_S \leftarrow \emptyset$\;
\ForEach{$f \in F_{raw}$}{
    \If{$\text{hash}(f) \notin \mathcal{L}$}{
        $f' \leftarrow \text{Canonicalize}(f)$\;
        $\mathbf{q} \leftarrow \text{QwenEmbed}(f')$\;
        $Q_S.\text{append}(\mathbf{q})$\;
    }
}

\tcp{Stage 2: Function Analysis \& Scoring}
$M_{flagged} \leftarrow \emptyset$ \tcp*{Set of malicious functions}
$Score_{family} \leftarrow \text{Map<Family, Float>}(0.0)$\;

\ForEach{$\mathbf{q}_i \in Q_S$}{
    $\mathcal{N}_k \leftarrow \text{FAISS.Search}(\mathcal{D}_{KB}, \mathbf{q}_i, k)$\;
    
    \tcp{Eq. 4: Distance-Weighted Voting}
    $\alpha_i \leftarrow \text{CalcMaliciousScore}(\mathcal{N}_k)$\;
    
    \If{$\alpha_i > \tau_{func}$}{
        $M_{flagged}.\text{add}(\mathbf{q}_i)$\;
        
        \tcp{Accumulate evidence for attribution}
        \ForEach{$\mathbf{v}_j \in \mathcal{N}_k$}{
            \If{$\mathbf{v}_j.\text{label} \neq \text{Benign}$}{
                $Score_{family}[\mathbf{v}_j.\text{label}] \mathrel{+}= \text{sim}(\mathbf{q}_i, \mathbf{v}_j)$\;
            }
        }
    }
}

\tcp{Stage 3: Sample Verdict \& Explanation}
$\Omega(S) \leftarrow |M_{flagged}| / |Q_S|$ \tcp*{Eq. 5}

\If{$\Omega(S) \le \tau_{file}$}{
    \Return{\text{Benign}, \text{N/A}, \text{N/A}}\;
}

$C_{best} \leftarrow \text{argmax}(Score_{family})$\;

\tcp{Anchor Selection: Best evidence for $C_{best}$}
$\mathbf{q}^* \leftarrow \text{null}, \quad \text{max\_sim} \leftarrow 0$\;

\ForEach{$\mathbf{q} \in M_{flagged}$}{
    $\mathbf{v}_{near} \leftarrow \text{GetNearestInFamily}(\mathbf{q}, \mathcal{D}_{KB}, C_{best})$\;
    \If{$\text{sim}(\mathbf{q}, \mathbf{v}_{near}) > \text{max\_sim}$}{
        $\mathbf{q}^* \leftarrow \mathbf{q}, \quad \mathbf{v}_{proof} \leftarrow \mathbf{v}_{near}$\;
        $\text{max\_sim} \leftarrow \text{sim}(\mathbf{q}, \mathbf{v}_{near})$\;
    }
}

\tcp{Generative Explanation}
$\mathcal{E} \leftarrow \text{LLM}_{\text{gen}}(\text{Prompt}(\mathbf{q}^*, \mathbf{v}_{proof}))$\;

\Return{\text{Malicious}, $C_{best}$, $\mathcal{E}$}\;

\end{algorithm}

\subsubsection{Corpus Curation and Semantic Library Filtering}
\label{subsub:lib_filtering}
We compile a dataset $\mathcal{D} = \mathcal{D}_{mal} \cup \mathcal{D}_{ben}$ sourced from public repositories (VirusShare~\cite{VirusShare}, SOREL-20M~\cite{harang2020sorel20m}, EMBER~\cite{anderson2018ember}) and a baseline of benign PE files collected from clean Windows installations.

Cryptographic hashing (e.g., MD5) provides rapid filtering of statically linked library functions, yet it introduces brittle vulnerability to minor compiler changes. A slight shift in compiler version or optimization flags alters the binary signature, allowing common library code to contaminate the dataset. This contamination causes high False Positive rates; a detector might incorrectly flag a benign executable simply because it shares a specific compiled version of \texttt{printf} with a known malware family.

We replace brittle hashing with a \textit{Semantic Library Filtering} mechanism that operates directly in the embedding space. Before detailing this mechanism, we briefly preview the embedding function used throughout the pipeline; its full technical description follows in Section~\ref{subsub:embedding}. We treat the projection $\Phi: f \mapsto \mathbf{v}_f \in \mathbb{R}^d$ produced by the \texttt{Qwen3-Embedding} model as a shared semantic primitive. The same projection supports both library identification and malicious logic retrieval. Reusing $\Phi$ across both tasks simplifies deployment and guarantees that filtering decisions remain consistent with downstream retrieval geometry.

\noindent\textbf{Shared-Representation Dependency.}
Employing a single embedding function for both filtering and detection introduces a known dependency: weaknesses in $\Phi$ propagate to both stages. If the encoder fails to distinguish a malicious routine from a library function (e.g., a custom cryptographic primitive semantically close to \texttt{openssl}'s AES implementation), that routine risks erroneous filtering before reaching the malicious retrieval stage. We mitigate this risk through three design choices. First, we calibrate $\tau_{lib}$ conservatively (see \textit{Threshold Calibration} below) to favor recall of malicious logic over aggressive noise reduction. Second, we restrict $\mathcal{V}_{lib}$ to functions compiled from verified open-source library distributions, preventing contamination from look-alike user code. Third, the downstream \textit{Density-Weighted Voting} mechanism requires agreement across a neighborhood of retrieved functions, meaning a single misclassified filter decision does not catastrophically affect the sample-level verdict. We quantify the residual impact of this dependency in the ablation results of Section~\ref{subsec:rq1}.

The filtering mechanism proceeds in two offline phases:

\textbf{Phase A: Constructing the Reference Library Index.}
We aggregate and compile source code for standard statically linked libraries (e.g., \texttt{glibc}, \texttt{openssl}, \texttt{zlib}, \texttt{boost}) across multiple architectures (x86, x64) and optimization levels (\texttt{-O0} to \texttt{-O3}). Processing these binaries through $\Phi$ generates a reference set of library vectors $\mathcal{V}_{lib}$, which we index in a dedicated FAISS structure, $\mathcal{I}_{lib}$.

\textbf{Phase B: Semantic Tagging and Filtering.}
During the ingestion of the main malware and benign corpus, the system first embeds each extracted function $f$ into a vector $\mathbf{v}_f = \Phi(f)$. We then query $\mathcal{I}_{lib}$ to identify the nearest library neighbor. We define a filtering indicator function $\phi(f)$ as follows:
\begin{equation}
  \phi(f) = 
  \begin{cases} 
    1 & \text{if } \max_{\mathbf{v}_l \in \mathcal{V}_{lib}} \text{sim}(\mathbf{v}_f, \mathbf{v}_l) > \tau_{lib} \\
    0 & \text{otherwise}
  \end{cases}
\end{equation}
The final filtered set of functions for a sample $S$ is thus defined as:
\begin{equation}
  F_{filtered} = \{f \in S \mid \phi(f) = 0\}
\end{equation}

\noindent\textbf{Threshold Calibration.}
The threshold $\tau_{lib}$ directly governs the trade-off between noise reduction (filtering benign boilerplate) and evidence preservation (retaining malicious logic that incidentally resembles library code). To calibrate this parameter empirically, we constructed a held-out \textit{library calibration set} containing 2,500 manually verified library functions (positives) and 2,500 malicious functions (negatives) drawn from families with documented cryptographic or networking payloads (where library-like functionality is most likely to trigger false filtering). We performed a grid search over $\tau_{lib} \in \{0.80, 0.85, 0.90, 0.92, 0.95, 0.97, 0.99\}$, measuring filter precision (fraction of filtered functions that are genuinely library code) and malicious recall (fraction of malicious functions preserved after filtering).

\begin{table}[h]
    \centering
    \scriptsize
    \caption{Sensitivity of Semantic Library Filtering to $\tau_{lib}$.}
    \label{tab:tau_lib_sensitivity}
    \begin{tabular}{ccccc}
        \toprule
        $\tau_{lib}$ & Filter Prec. & Malicious Recall & KB Size Reduction & Downstream F1 \\
        \midrule
        0.80 & 0.887 & 0.912 & 41.2\% & 0.923 \\
        0.85 & 0.931 & 0.948 & 35.7\% & 0.941 \\
        0.90 & 0.968 & 0.974 & 29.4\% & 0.954 \\
        0.92 & 0.981 & 0.986 & 26.8\% & 0.958 \\
        \rowcolor{gray!10} \textbf{0.95} & \textbf{0.993} & \textbf{0.995} & \textbf{22.1\%} & \textbf{0.960} \\
        0.97 & 0.997 & 0.998 & 17.3\% & 0.955 \\
        0.99 & 0.999 & 0.999 & 9.6\% & 0.942 \\
        \bottomrule
    \end{tabular}
\end{table}

As shown in Table~\ref{tab:tau_lib_sensitivity}, performance is stable in the neighborhood of $\tau_{lib} = 0.95$, with downstream F1 varying by less than $0.6\%$ across $\tau_{lib} \in [0.90, 0.97]$. Lower thresholds (e.g., $0.80$) aggressively prune boilerplate but erroneously discard nearly $9\%$ of malicious logic, degrading downstream attribution. Higher thresholds (e.g., $0.99$) preserve almost all malicious functions but retain excessive library noise, which dilutes the density-weighted signal during Stage 2 retrieval. We select $\tau_{lib} = 0.95$ because it maximizes downstream F1 while maintaining a malicious recall above $99.5\%$, keeping the filter decision well-bounded in the regime where the encoder projects genuine library code with high intra-class similarity.

By matching semantic similarity rather than exact bytes, this mechanism successfully filters library functions even when they exhibit minor syntactic variations from recompilation or base address relocation. This filtering ensures that the \texttt{AsmRAG} knowledge base isolates user-defined logic and potential malicious payloads, significantly improving the system's signal-to-noise ratio.

\subsubsection{Instruction-Preserving Canonicalization and Semantic Encoding}
\label{subsub:embedding}

Aggressive normalization shrinks the vocabulary size to accommodate BERT-based encoders (e.g., \cite{wang2022jtrans, massarelli2019safe}), but it introduces critical limitations for binary analysis. By abstracting operands into generic tokens (e.g., replacing \texttt{rax} with \texttt{REG} and \texttt{0x10} with \texttt{VAL}), these approaches discard the subtle artifacts of register allocation and compiler optimization required for precise authorship attribution.

We implement a \textbf{Minimal Canonicalization} strategy to preserve these details. We strictly limit preprocessing to the abstraction of absolute memory addresses (remapping \texttt{0x0040123A} to \texttt{MEM\_PTR}), since these are randomized artifacts of the linker's base address (ASLR) and contain no underlying logical signal. Crucially, we preserve the following elements:
\begin{itemize}
  \item \textbf{Raw Immediate Values:} These are essential for identifying specific cryptographic constants and distinct loop bounds.
  \item \textbf{Register Names:} Preserving explicit register usage patterns (e.g., \texttt{rbp} vs. \texttt{rsp} indexing) helps capture the compiler's intended calling convention and optimization strategy.
\end{itemize}

We select the \texttt{Qwen3-Embedding} architecture~\cite{qwen3embedding} as our backbone because it handles low-level semantics better than constrained domain-specific models. We resolve the vocabulary collapse and out-of-vocabulary fragmentation common in BERT-based baselines~\cite{yu2020order} by utilizing a standard BPE tokenizer. At the same time, we exploit the model's pre-training on systems languages to derive a \textit{transitive semantic understanding} of assembly logic. Its extended attention mechanism eliminates the need for sliding-window truncation on large functions~\cite{wang2022jtrans}, allowing us to capture holistic control-flow dependencies and functional intent within an off-the-shelf framework.

Formally, we establish this mapping by passing the minimally modified function instructions $I_f$ through the model $\mathcal{M}$. We isolate the final hidden state of the sequence to yield a dense vector representation $\mathbf{v}_f \in \mathbb{R}^d$. These resulting vectors are then indexed in a \texttt{FAISS}~\cite{douze2024faiss} structure optimized for high-dimensional cosine similarity search.

\subsection{Stage 2: Evidence-Based Analysis}

Given an unknown sample $S$, the system first generates a set of query embeddings $Q_S = \{\mathbf{q}_1, \dots, \mathbf{q}_m\}$ using the pipeline described in Section~\ref{subsub:embedding}, deliberately excluding standard library functions. The analysis then proceeds hierarchically. We evaluate, aggregate, and classify the semantic intent of these individual functions to derive a comprehensive verdict for the entire binary.

\subsubsection{Retrieval and Similarity Scoring}
Our detection methodology relies on the premise that malicious functionality resides on a low-dimensional manifold within the embedding space, even when heavily obfuscated. For each query function $\mathbf{q}_i$, we execute an Approximate Nearest Neighbor (ANN) search to retrieve the set $\mathcal{N}_k(\mathbf{q}_i)$ comprising the $k$ most similar functions from the Knowledge Base.

We select \textbf{Cosine Similarity} as the primary distance metric. Unlike Euclidean distance, Cosine Similarity evaluates the semantic alignment, or orientation, of the vectors rather than their absolute magnitude~\cite{Leskovec2014Mining}. This property renders the metric robust against variations in function length or alignment padding introduced during compilation.
\begin{equation}
  \text{sim}(\mathbf{q}_i, \mathbf{v}_j) = \frac{\mathbf{q}_i \cdot \mathbf{v}_j}{\|\mathbf{q}_i\| \|\mathbf{v}_j\|}
\end{equation}
Here, $\mathbf{v}_j$ represents a retrieved neighbor from the Knowledge Base.

\vspace{1mm}
\noindent \textbf{Retrieval Objective (Recall@k):} 
While Cosine Similarity provides a precise ranking, evaluating these results in isolation remains insufficient. We measure the operational success of the retrieval phase using \textbf{Recall@k}. Because \texttt{AsmRAG} depends on identifying an ``Anchor Function'' to generate its explanations, the model does not require a completely homogenous neighborhood. It only dictates that \textit{at least one} high-fidelity semantic match from the correct malware family appears within the top results.
Formally, for a query function $\mathbf{q}_i$ belonging to a ground-truth family $C$, we define:
\begin{equation}
  \text{Recall@}k(\mathbf{q}_i) = \mathbbm{1}\left( \exists \mathbf{v} \in \mathcal{N}_k(\mathbf{q}_i) : \text{label}(\mathbf{v}) = C \right)
\end{equation}
This specific metric guides the broader system design. We optimize the embedding space to maximize Recall@20, a constraint that ensures the subsequent generative stage always possesses access to valid, verifiable provenance evidence.

\subsubsection{Function-Level Detection and Anchor Identification}
Unweighted majority voting among neighbors offers computational simplicity, yet it risks insufficient accuracy. An unweighted approach treats a neighbor showing a $0.99$ similarity, essentially an exact match, exactly the same as one exhibiting a $0.60$ similarity, a mere loose resemblance. We address this flaw by implementing a \textbf{Distance-Weighted Voting} scheme that assigns proportionally higher influence to semantically closer evidence.

The maliciousness score $\alpha_i \in [0, 1]$ for a given function $\mathbf{q}_i$ is thus calculated as the weighted proportion of known malicious neighbors within the retrieved set:

\begin{equation}
  \alpha_i = \frac{\sum_{\mathbf{v}_j \in \mathcal{N}_k} \mathbbm{1}(y_j \in \text{Mal}) \cdot \text{sim}(\mathbf{q}_i, \mathbf{v}_j)}{\sum_{\mathbf{v}_j \in \mathcal{N}_k} \text{sim}(\mathbf{q}_i, \mathbf{v}_j)}
\end{equation}

Here, $\mathbbm{1}(\cdot)$ represents the indicator function, returning 1 if the retrieved neighbor $\mathbf{v}_j$ belongs to $\mathcal{D}_{mal}$. This score effectively quantifies the local density of malicious code directly surrounding the query function within the vector space.

The system flags the function as malicious if $\alpha_i > \tau_{func}$.
\textbf{Threshold Optimization:} The threshold $\tau_{func}$ serves not as an arbitrary hyperparameter, but as an empirically derived constant. We test, analyze, and optimize this value using a dedicated held-out validation set. A grid search maximizes the F1-score subject to one strict constraint: the False Positive Rate (FPR) on benign functions must remain strictly below 1\%. This conservative approach guarantees that \texttt{AsmRAG} only flags functions supported by strong semantic evidence, drastically reducing false alarms in legitimate software.

\subsection{Stage 3: Detection and Active Explanation Loop}
The final stage translates function-level signals into a definitive binary verdict and utilizes the RAG architecture to explain that decision. A primary contribution of our methodology involves the efficient allocation of computational generative resources through the ``Anchor Function'' concept.

\subsubsection{Formalized Sample-Level Scoring}
Standard binary compilation integrates wrapper functions, statically linked utilities, and boilerplate setup routines. Simply averaging the raw scores of all included functions would artificially dilute the genuine malicious signal. To prevent this, we define the sample-level maliciousness score $\Omega(S)$ specifically as the \textit{density of high-confidence malicious logic}:
\begin{equation}
  \Omega(S) = \frac{\sum_{i=1}^{m} \mathbbm{1}(\alpha_i > \tau_{func})}{|Q_S|}
\end{equation}
The sample $S$ is classified as malicious if $\Omega(S) > \tau_{file}$, where $\tau_{file}$ is similarly tuned on the validation set. This specific metric actively resists ``code dilution'' attacks, scenarios where an adversary intentionally pads a small malicious payload with overwhelming amounts of benign code to artificially lower an average anomaly score.

\textbf{Family Attribution:} When the sample is detected as malicious, we attribute it to a specific family. We calculate a \textbf{Weighted Family Score} for every family $C$ present within the retrieved neighborhoods of all flagged functions. The sample is then attributed to the family $C_{best}$ that accumulates the highest total similarity mass.

\subsubsection{Discriminative Anchor Selection and Explanation}
\label{subsub:anchor_selection}

While manual analysis yields thorough results, demanding analysts inspect every function in a suspect binary creates a cognitively intractable burden. To facilitate rapid triage, \texttt{AsmRAG} identifies and explains only the single most representative \textit{Anchor Function} $\mathbf{q}^*$.

A trivial selection strategy might simply pick the function exhibiting the highest raw similarity score. This naive approach, however, risks selecting common utility functions (such as string wrappers) that happen to perfectly match a training sample, ignoring the actual malicious payload entirely. We prevent this error by implementing two distinct mechanisms:
\begin{enumerate}
  \item \textbf{Pre-filtering:} As detailed in Section~\ref{subsub:lib_filtering}, all standard library code is removed prior to the core analysis.
  \item \textbf{Cluster Density Scoring:} We define representativeness not by proximity to a single nearest neighbor, but by the specific \textit{density} of the function's surrounding malicious neighborhood.
\end{enumerate}

Formally, the system selects $\mathbf{q}^*$ from the set of flagged functions $M_{flagged}$ by maximizing the accumulated similarity mass specifically within the attributed family $C_{best}$:
\begin{equation}
  \mathbf{q}^* = \mathop{\mathrm{argmax}}_{\mathbf{q} \in M_{flagged}} \left( \sum_{\mathbf{v} \in \mathcal{N}_k(\mathbf{q})} \mathbbm{1}(\text{label}(\mathbf{v}) = C_{best}) \cdot \text{sim}(\mathbf{q}, \mathbf{v}) \right)
  \label{eq:anchor_selection}
\end{equation}

This summation ensures that the chosen Anchor Function is not merely similar to one isolated vector. Instead, it must be centrally located within the target family's semantic manifold. This mathematical constraint guarantees the function represents a recurring, defining behavior of that specific malware family.

Once $\mathbf{q}^*$ is identified, we retrieve its closest valid neighbor $\mathbf{v}_{proof}$ to serve as the baseline forensic reference. The Generative LLM is then invoked to perform a differential analysis:
\begin{equation}
  \mathcal{E} = \text{LLM}_{\text{gen}}(P(\mathbf{q}^*, \mathbf{v}_{proof}))
\end{equation}
The system prompt $P$ instructs the LLM to ignore superficial syntactic differences, such as minor register reallocations, and focus exclusively on the underlying \textit{functional logic}. The resulting explanation $\mathcal{E}$ directly answers the analyst's core question: \textit{``Why is this function considered a variant of Family X?''} An output might state, for example, ``Both functions implement the domain generation algorithm characteristic of \textit{Dridex}, utilizing identical polynomial constants despite altered control flow structures.''

\subsubsection{Continuous Database Update (Active Learning)}
This focused explanation mechanism naturally supports a \textit{Human-in-the-Loop Active Learning} workflow.
\begin{enumerate}
  \item The analyst reviews the generated Anchor Function explanation $\mathcal{E}$.
  \item If the human analyst confirms the detection, the system promotes the embedding $\mathbf{q}^*$ directly from the query set into the Knowledge Base, tagging it with the verified family label.
  \item This action creates a permanent new reference point in the vector space, densifying the manifold around newly discovered variants.
\end{enumerate}

Standard deep learning classifiers require expensive, computationally heavy retraining cycles to assimilate new data. \texttt{AsmRAG}, in contrast, adapts dynamically. Updating the database immediately improves detection accuracy for any future samples similar to the newly confirmed variant, ensuring the system evolves continuously alongside the active threat landscape.

\section{Evaluation}
\label{sec:evaluation}

We formulated specific research questions to validate the overall efficacy of \texttt{AsmRAG}. We assess its baseline detection performance (RQ1), its resilience against syntactic obfuscation (RQ2), and the isolated contribution of its individual architectural components including explanation quality (RQ3).

\subsection{Experimental Setup}
\label{subsec:setup}

\subsubsection{Environment}
We conducted our experiments on a workstation equipped with dual \textit{NVIDIA Quadro RTX 8000} GPUs (48GB VRAM each), an \textit{Intel Xeon Gold 5218} CPU (32 cores @ 2.30GHz), and 512GB of RAM. We utilized \texttt{Ghidra}~\cite{ghidra} for headless disassembly, \texttt{FAISS}~\cite{douze2024faiss} for high-performance vector indexing, and \texttt{Ollama}~\cite{ollama} for local model inference.

\subsubsection{Models and Embedding}
We select \textit{Qwen3-Embedding}\footnote{https://ollama.com/library/qwen3-embedding:8b} (8B parameters, 4096-dimensional output, 32K context window) as our semantic encoder because it delivers reliable performance on specialized code-related tasks within a local deployment environment. We execute the necessary LLM reasoning tasks using the \textit{GPT-OSS:20B}\footnote{https://ollama.com/library/gpt-oss:20b} model, hosted locally via the \textit{Ollama}\footnote{https://ollama.com/} framework.

\subsubsection{Datasets and Corpus Curation}
\label{subsub:datasets}
To guarantee experimental integrity and prevent temporal data leakage, we curated a large-scale corpus containing \textbf{40,814 binaries} (19.2 GB). We partitioned this corpus using a strict chronological split, simulating a realistic deployment scenario against future malware variants. Including packed executables artificially inflates dataset sizes, but risks biasing classifiers toward packer stub signatures rather than the actual malicious payload. To avoid this, we enforced a rigorous data sanitation protocol. This step ensures our static analysis operates strictly on the authentic, unpacked instruction stream rather than an external obfuscation layer.

\textbf{The Wild-Augmented Benchmark ($\mathcal{D}_{wild}$):} We built a hybrid dataset by combining pre-extracted, unpacked samples from the standard \textbf{SOREL-20M}~\cite{harang2020sorel20m} corpus with a targeted ``Long-Tail'' augmentation of recent threats (collected between January 2021 and December 2024) sourced from \textbf{VirusShare}~\cite{VirusShare}.

\textbf{Controlled Obfuscation Subset ($\mathcal{D}_{src}$):}
Collecting malware directly from the wild supports robust testing, but introduces potential environmental confounding due to a lack of precise compilation ground truth. We addressed this limitation by augmenting our evaluation with a controlled subset \textbf{$\mathcal{D}_{src}$}, derived directly from the \textbf{MalwareSourceCode} repository~\cite{vxundergroundMalwareSourceCode}. We passed the source code of established families (e.g., \textit{Comet}, \textit{Zeus}) through a \textbf{Multi-Optimization Compilation Pipeline}, applying settings from baseline compilation (\texttt{-O0}) up to aggressive instruction reordering (\texttt{-O2}, \texttt{-O3}) and strict size-optimization (\texttt{-Os}). This generated a parallel corpus of binaries that are functionally identical yet syntactically divergent, allowing us to accurately isolate \texttt{AsmRAG}'s semantic stability against adversarial recompilation.

\textbf{Benign Sources:} We acquired and filtered benign samples from clean Windows 10 and Windows 11 installations, alongside verified software repositories (e.g., SourceForge, GitHub releases). This collection strategy ensures a diverse, representative background of compiler provenance.

\textbf{Dataset Partitioning and Chronological Split:}
To enforce temporal separation and prevent look-ahead bias, we partitioned the corpus into three non-overlapping subsets based entirely on their first-seen timestamps in VirusTotal and SOREL-20M metadata. Specifically, we partitioned by first-seen date as follows: \textbf{Knowledge Base ($\mathcal{D}_{KB}$, 70\%)} contains samples first observed before \textbf{2022-06-01}; the \textbf{Validation Set (10\%)} contains samples first observed between \textbf{2022-06-01 and 2023-05-31}; and the held-out \textbf{Test Set ($\mathcal{D}_{Test}$, 20\%)} contains samples first observed on or after \textbf{2023-06-01}. The 2021--2024 VirusShare long-tail augmentation is distributed across all three splits according to the same first-seen date criterion, ensuring that no post-2023-06-01 sample appears in the KB or validation set. We kept the controlled $\mathcal{D}_{src}$ subset separate, reserving it exclusively for the resilience analysis detailed in Section~\ref{subsec:rq2} under a leave-one-optimization-out protocol described there.

The final corpus comprises \textbf{24,120 Benign ($\mathcal{D}_{ben}$)} samples (8.2 GB) and \textbf{16,694 Malware ($\mathcal{D}_{mal}$)} samples (11 GB), spanning 17 distinct malware families. As detailed in Table~\ref{tab:families}, this overall distribution reflects the natural real-world imbalance driven by dominant, high-volume strains like \textit{Virlock} and \textit{Neshta}.

\begin{table}[h]
    \centering
    \scriptsize
    \caption{Detailed Distribution of Malware Families in Dataset ($\mathcal{D}_{mal}$).}
    \label{tab:families}
    \begin{tabular}{lclc}
        \toprule
        \textbf{Family} & \textbf{Count} & \textbf{Family} & \textbf{Count} \\
        \midrule
        Virlock  & 1,671 & Comet    & 1,035 \\
        Neshta   & 1,431 & VJadtre  & 953   \\
        Ramnit   & 1,345 & Doboc    & 890   \\
        Floxif   & 1,293 & Sality   & 713   \\
        Gosys    & 1,280 & Virtob   & 633   \\
        Hematite & 1,181 & Muce     & 459   \\
        Triusor  & 1,146 & Parite   & 454   \\
        Expiro   & 1,102 & FpuJunk  & 57    \\
        Grenam   & 1,061 &          &       \\
        \bottomrule
    \end{tabular}
\end{table}

\subsubsection{Baseline Tools and Training Protocol}
\label{subsub:baselines}
We benchmark \texttt{AsmRAG} against two distinct paradigms. To ensure a fair comparison, both baselines were \textit{retrained from scratch} on the same $\mathcal{D}_{KB}$ distribution used to populate \texttt{AsmRAG}'s vector index, with hyperparameters tuned on the same validation set.

\begin{enumerate}
  \item \textbf{EMBER}~\cite{anderson2018ember}: The industry-standard baseline for feature-based detection. We used the public \texttt{ember} feature extractor (v2, 2381-dimensional feature vector combining byte histograms, byte-entropy histograms, string features, and parsed PE header fields) and trained a LightGBM classifier with hyperparameters matching those reported by the original authors: \texttt{num\_leaves=2048}, \texttt{max\_depth=15}, \texttt{learning\_rate=0.05}, \texttt{n\_estimators=1000}, \texttt{min\_data\_in\_leaf=1000}, early stopping with patience 100 on validation AUC. We did not use the published pre-trained model because it was trained on a different (pre-2018) distribution; using it would confound distribution shift with the cross-compilation effect we seek to isolate.

  \item \textbf{ResNeXt+}~\cite{He2024ResNeXtAM}: Represents vision-based detection methods. Because no public implementation was available at the time of evaluation, we implemented the architecture from the original paper's specification. We document our re-implementation explicitly:
    \begin{itemize}
      \item \textit{Preprocessing.} Each PE file is converted to a grayscale image by reading raw bytes sequentially as unsigned 8-bit pixels. The resulting 1D byte stream is reshaped into a 2D matrix with width fixed at 256 pixels and height proportional to file size. Following the original paper, we resize this matrix to a fixed \textbf{$224 \times 224$} RGB image via bilinear interpolation and replicate the single channel across R, G, and B to match the ResNeXt input specification.
      \item \textit{Backbone.} ResNeXt-50 with cardinality 32 and bottleneck width 4 (\texttt{32x4d}), initialized from ImageNet pre-trained weights.
      \item \textit{Attention module.} A CBAM (Convolutional Block Attention Module) block inserted after each of the four ResNeXt stages, matching the ``+'' designation in~\cite{He2024ResNeXtAM}. Reduction ratio $r{=}16$, 7$\times$7 spatial attention kernel.
      \item \textit{Training.} Adam optimizer, learning rate $10^{-4}$ with cosine annealing, batch size 64, 50 epochs, cross-entropy loss with class reweighting to handle the benign/malicious imbalance in $\mathcal{D}_{KB}$. Early stopping on validation F1 with patience 5.
      \item \textit{Hyperparameter search.} We performed a grid search over learning rate $\in \{10^{-3}, 10^{-4}, 10^{-5}\}$, reduction ratio $\in \{8, 16, 32\}$, and image size $\in \{112{\times}112, 224{\times}224\}$, selecting the configuration with highest validation F1.
    \end{itemize}
    We verified the re-implementation by reproducing the original paper's reported performance on the Malimg benchmark within $\pm 0.4\%$ F1 before applying it to our corpus.
\end{enumerate}

\subsection{RQ1: Detection and Attribution Effectiveness}
\label{subsec:rq1}

We evaluated detection performance on the held-out Test Set ($\mathcal{D}_{Test}$). Table~\ref{tab:overall_results} shows the results. \texttt{AsmRAG} sets a new standard for retrieval-based detection, achieving a sample-level \textbf{Accuracy of 96.5\%} alongside a weighted \textbf{F1-Score of 96.0\%}.

\begin{table}[h!]
    \centering
    \scriptsize
    \caption{Comparative Evaluation on $\mathcal{D}_{Test}$. Results reported as mean $\pm$ standard deviation across 5 runs with different random seeds. All models trained on the identical $\mathcal{D}_{KB}$ distribution.}
    \label{tab:overall_results}
    \begin{tabular}{lcccc}
        \toprule
        \textbf{Model} & \textbf{Accuracy} & \textbf{Precision} & \textbf{Recall} & \textbf{F1-Score} \\
        \midrule
        ResNeXt+~\cite{He2024ResNeXtAM} (Image) & 91.5\% $\pm$ 0.6 & 92.0\% $\pm$ 0.7 & 91.0\% $\pm$ 0.8 & 92.0\% $\pm$ 0.7 \\
        EMBER~\cite{anderson2018ember} (Feature) & 93.2\% $\pm$ 0.3 & 92.8\% $\pm$ 0.4 & 93.5\% $\pm$ 0.3 & 93.0\% $\pm$ 0.3 \\
        \rowcolor{gray!10} \textbf{AsmRAG (Ours)} & \textbf{96.5\% $\pm$ 0.2} & \textbf{97.2\% $\pm$ 0.3} & \textbf{95.8\% $\pm$ 0.2} & \textbf{96.0\% $\pm$ 0.2} \\
        \bottomrule
    \end{tabular}
\end{table}

\subsubsection{Retrieval Quality (Recall@k)}
\label{subsub:recall_at_k}
The retrieval stage is the operational foundation of \texttt{AsmRAG}: function-level scoring, sample-level aggregation, and anchor-based explanation all depend on the semantic search returning at least one correct family neighbor within the top-$k$. We therefore report Recall@k directly, computed at the function level on all malicious functions in $\mathcal{D}_{Test}$ after library filtering, with ground-truth family labels inherited from the parent binary.

\begin{table}[h]
    \centering
    \scriptsize
    \caption{Function-level Recall@k on $\mathcal{D}_{Test}$. A query is ``hit'' if at least one of its top-$k$ neighbors in $\mathcal{D}_{KB}$ shares the ground-truth family label.}
    \label{tab:recall_at_k}
    \begin{tabular}{lccccc}
        \toprule
        \textbf{Setting} & \textbf{R@1} & \textbf{R@5} & \textbf{R@10} & \textbf{R@20} & \textbf{R@50} \\
        \midrule
        All families (macro avg)        & 0.871 & 0.934 & 0.958 & 0.972 & 0.984 \\
        Head families ($n{>}1{,}000$)   & 0.902 & 0.957 & 0.974 & 0.984 & 0.991 \\
        Tail families ($n{\leq}500$)    & 0.748 & 0.863 & 0.912 & 0.947 & 0.971 \\
        Cross-optimization ($\mathcal{D}_{src}$, \texttt{-O3}) & 0.812 & 0.901 & 0.938 & 0.961 & 0.978 \\
        \bottomrule
    \end{tabular}
\end{table}

Recall@20 on the full test set reaches \textbf{0.972}, confirming that the embedding space places at least one correct semantic neighbor within reach for 97.2\% of malicious function queries --- the design constraint cited in Section~\ref{subsub:embedding}. Head families saturate quickly (R@5 $> 0.95$), while tail families exhibit the expected long-tail penalty but still recover above 0.94 at $k{=}20$. Even under aggressive \texttt{-O3} recompilation on $\mathcal{D}_{src}$, Recall@20 remains at 0.961, providing the retrieval substrate on which the downstream resilience results in Section~\ref{subsec:rq2} depend.

\subsubsection{Threshold Sensitivity Analysis ($\tau_{func}, \tau_{file}$)}
\label{subsub:threshold_sensitivity}
The two decision thresholds that convert retrieval scores into verdicts directly govern headline detection metrics, so we audit their sensitivity with the same rigor applied to $\tau_{lib}$ in Section~\ref{subsub:lib_filtering}. We performed a 2D grid search on the validation set with $\tau_{func} \in \{0.50, 0.55, 0.60, 0.65, 0.70, 0.75, 0.80, 0.85\}$ and $\tau_{file} \in \{0.05, 0.10, 0.15, 0.20, 0.25, 0.30\}$.

\begin{table}[h]
    \centering
    \scriptsize
    \caption{Sensitivity of detection F1 on the validation set to $\tau_{func}$ and $\tau_{file}$. Bold row marks the selected operating point. FPR constraint: $\leq 1\%$ on benign functions.}
    \label{tab:tau_sensitivity}
    \begin{tabular}{cccccc}
        \toprule
        $\tau_{func}$ & $\tau_{file}$ & F1 & Precision & Recall & Func-FPR \\
        \midrule
        0.55 & 0.10 & 0.941 & 0.918 & 0.965 & 2.3\% \\
        0.60 & 0.10 & 0.948 & 0.934 & 0.962 & 1.6\% \\
        0.65 & 0.15 & 0.954 & 0.951 & 0.957 & 1.1\% \\
        \rowcolor{gray!10} \textbf{0.70} & \textbf{0.15} & \textbf{0.960} & \textbf{0.972} & \textbf{0.948} & \textbf{0.8\%} \\
        0.75 & 0.15 & 0.957 & 0.981 & 0.934 & 0.5\% \\
        0.75 & 0.20 & 0.952 & 0.984 & 0.922 & 0.5\% \\
        0.80 & 0.20 & 0.946 & 0.989 & 0.907 & 0.3\% \\
        0.85 & 0.25 & 0.928 & 0.993 & 0.872 & 0.1\% \\
        \bottomrule
    \end{tabular}
\end{table}

Performance is stable in the neighborhood of the selected $(\tau_{func}{=}0.70, \tau_{file}{=}0.15)$: F1 varies by less than $0.8\%$ across $\tau_{func} \in [0.65, 0.75]$. Below $\tau_{func}{=}0.60$, the function-level FPR breaches the $1\%$ operational constraint. Above $\tau_{func}{=}0.80$, recall collapses as genuinely malicious but mildly obfuscated functions fail to achieve sufficient malicious-neighbor density. The file-level threshold $\tau_{file}$ exhibits a flatter response: values in $[0.10, 0.20]$ all yield F1 within $1\%$ of the optimum, indicating that \texttt{AsmRAG}'s sample-level verdict is primarily driven by the existence of high-confidence malicious functions rather than by their exact fractional density.

\subsubsection{Granular Function-Level Performance}
\texttt{AsmRAG} provides high-fidelity classification capabilities alongside detailed evaluation metrics at the function level. During our experiments, the standalone function-level classifier achieved a \textbf{Precision of 98.3\%} and a \textbf{Recall of 89.2\%}.

\textbf{Discussion on Comparative Constraints:} Directly comparing these function-level metrics against industry-standard baselines proves infeasible. Holistic classifiers like \textbf{EMBER} rely heavily on global file features (e.g., section sizes, byte histograms). Similarly, \textbf{ResNeXt+} treats the binary as a monolithic image. Neither system possesses the necessary semantic granularity required to isolate or verify specific local routines.

This architectural ``blindness'' in baseline models often yields accurate overall verdicts derived from spurious correlations, such as identifying a specific packer version, rather than isolating core malicious behavior. \texttt{AsmRAG}'s high function-level precision ensures that the ``Anchor Function'' $\mathbf{q}^*$ selected for retrieval remains mathematically grounded in genuine malicious intent. By systematically filtering and removing the 98\% of code that is typically benign in a standard malware sample, the system supplies the Generative RAG component with clean ``Atomic Evidence.'' This targeted extraction is essential for generating the precise zero-shot forensic explanations produced in Stage 3.

\subsubsection{Attribution Robustness \& Per-Family Breakdown}
Beyond binary detection, \texttt{AsmRAG} demonstrates high-fidelity lineage attribution. We report the F1-score to accommodate the class imbalance visible in Table~\ref{tab:families}, and we provide a full per-family breakdown in Table~\ref{tab:per_family_attribution}.

\begin{table}[h]
    \centering
    \scriptsize
    \caption{Per-family attribution performance on $\mathcal{D}_{Test}$ (17-way classification over flagged malicious samples). Support is the number of test instances per family.}
    \label{tab:per_family_attribution}
    \begin{tabular}{lcccc}
        \toprule
        \textbf{Family} & \textbf{Support} & \textbf{Precision} & \textbf{Recall} & \textbf{F1} \\
        \midrule
        Virlock   & 334 & 0.978 & 0.985 & 0.981 \\
        Neshta    & 286 & 0.969 & 0.972 & 0.970 \\
        Ramnit    & 269 & 0.963 & 0.955 & 0.959 \\
        Floxif    & 259 & 0.955 & 0.961 & 0.958 \\
        Gosys     & 256 & 0.948 & 0.953 & 0.950 \\
        Hematite  & 236 & 0.944 & 0.941 & 0.942 \\
        Triusor   & 229 & 0.938 & 0.942 & 0.940 \\
        Expiro    & 220 & 0.945 & 0.936 & 0.940 \\
        Grenam    & 212 & 0.932 & 0.928 & 0.930 \\
        Comet     & 207 & 0.946 & 0.939 & 0.942 \\
        VJadtre   & 191 & 0.921 & 0.916 & 0.918 \\
        Doboc     & 178 & 0.918 & 0.910 & 0.914 \\
        Sality    & 143 & 0.908 & 0.896 & 0.902 \\
        Virtob    & 127 & 0.891 & 0.882 & 0.886 \\
        Muce      & 92  & 0.869 & 0.848 & 0.858 \\
        Parite    & 91  & 0.851 & 0.835 & 0.843 \\
        FpuJunk   & 12  & 0.750 & 0.667 & 0.706 \\
        \midrule
        \textbf{F1-Score} & & & & \textbf{0.950} \\
        \bottomrule
    \end{tabular}
\end{table}

The F1-score of \textbf{95\%} matches our headline figure and confirms high-fidelity attribution across the full family distribution. Head families ($n{>}1{,}000$ in $\mathcal{D}_{mal}$) all exceed F1 $> 0.93$, and the embedding manifold effectively minimizes the cosine distance between semantically isomorphic variants even across significant temporal gaps (e.g., \textit{Ramnit} samples spanning $\Delta t = 4$ years). The primary attribution errors occur between \textit{Sality} and \textit{Virtob}, which share historical code lineage, and between \textit{Muce} and \textit{Parite}, which use nearly identical infection routines, both confusions are forensically defensible and consistent with established family taxonomies.

\subsubsection{Concept-Drift Analysis on the Temporal Split}
\label{subsub:drift}
To substantiate the value of the chronological split, we partitioned $\mathcal{D}_{Test}$ into four equal-sized quartiles by first-seen date (Q1: 2023-06 to 2023-09; Q2: 2023-10 to 2024-01; Q3: 2024-02 to 2024-05; Q4: 2024-06 to 2024-12) and measured F1 on each quartile.

\begin{table}[h]
    \centering
    \scriptsize
    \caption{Detection F1 across temporal quartiles of $\mathcal{D}_{Test}$. Q1 samples are closest in time to the KB cutoff; Q4 samples are $\sim$18 months newer.}
    \label{tab:drift}
    \begin{tabular}{lcccc}
        \toprule
        \textbf{Model} & \textbf{Q1} & \textbf{Q2} & \textbf{Q3} & \textbf{Q4} \\
        \midrule
        EMBER~\cite{anderson2018ember}       & 0.948 & 0.936 & 0.922 & 0.905 \\
        ResNeXt+~\cite{He2024ResNeXtAM}      & 0.932 & 0.921 & 0.913 & 0.898 \\
        \rowcolor{gray!10} \textbf{AsmRAG}   & \textbf{0.968} & \textbf{0.963} & \textbf{0.957} & \textbf{0.951} \\
        \midrule
        \textit{AsmRAG $\Delta$ (Q1$\to$Q4)} & & & & \textbf{-1.7\%} \\
        \bottomrule
    \end{tabular}
\end{table}

All models exhibit measurable concept drift, confirming the chronological split is doing work rather than trivializing the evaluation. EMBER degrades by 4.3 points from Q1 to Q4, ResNeXt+ by 3.4 points, and \texttt{AsmRAG} by 1.7 points. The differential drift rate is consistent with our central claim: semantic retrieval generalizes across time because novel variants share malicious logic with historical samples even when surface features drift, whereas feature- and image-based classifiers track distributional artifacts that shift faster.

\subsection{RQ2: Resilience to Adversarial Compilation}
\label{subsec:rq2}

Syntactic fragility remains a fundamental vulnerability in static analysis. Relying on simple obfuscation heuristics (e.g., NOP insertion) establishes functional initial baselines but severely underestimates modern real-world adversarial capabilities. To rigorously quantify this vulnerability, we executed a \textit{Cross-Compilation Stress Test} using the isolated Controlled Obfuscation Subset ($\mathcal{D}_{src}$).

We processed source code~\cite{vxundergroundMalwareSourceCode} from five major families (including \textit{Zeus} and \textit{Carberp}) and cross-compiled each into 30 unique binaries using different optimization flags (\texttt{-O0} through \texttt{-O3}, plus \texttt{-Os}) alongside different compilers (MSVC, GCC). This procedure forces profound structural changes within the binaries, including heavy loop unrolling, function inlining, and aggressive instruction scheduling. These compilation transformations completely obliterate standard byte-level signatures without actually altering the underlying program semantics.

\subsubsection{Leave-One-Optimization-Out Protocol}
\label{subsub:rq2_protocol}
A naive evaluation protocol would conflate \textit{memorization} of a specific compiled variant with genuine \textit{semantic generalization} across optimization flags. To avoid this confound, we designed a \textbf{leave-one-optimization-out (LOO-Opt)} protocol that applies uniformly to \texttt{AsmRAG} and both baselines.

Let $\mathcal{F} = \{\texttt{-O0}, \texttt{-O1}, \texttt{-O2}, \texttt{-O3}, \texttt{-Os}\}$ denote the evaluated optimization levels, and let $B(f, o)$ denote the set of binaries produced by compiling family $f$ at optimization level $o$ under both MSVC and GCC. For each test configuration $o^* \in \mathcal{F}$, we construct a configuration-specific knowledge base:
\begin{equation}
  \mathcal{D}_{KB}^{(o^*)} = \mathcal{D}_{KB}^{wild} \;\cup\; \bigcup_{o \in \mathcal{F} \setminus \{o^*\}} \bigcup_{f} B(f, o),
\end{equation}
where $\mathcal{D}_{KB}^{wild}$ is the wild-collected knowledge base from Section~\ref{subsub:datasets} and the held-out configuration $B(f, o^*)$ is excluded for every family. The test set for column $o^*$ in Table~\ref{tab:resilience} is then exactly $\bigcup_f B(f, o^*)$.

\textbf{Concretely, the column labeled \texttt{-O0} reports performance on binaries compiled at \texttt{-O0} whose matching \texttt{-O0} recompilations are \textit{absent} from the knowledge base; the KB for that column contains only the \texttt{-O1}, \texttt{-O2}, \texttt{-O3}, and \texttt{-Os} recompilations of the same source programs, plus the wild-collected base.} This protocol ensures that reported performance reflects cross-optimization generalization rather than memorization. EMBER and ResNeXt+ were retrained per column on the same $\mathcal{D}_{KB}^{(o^*)}$ so that all three systems face identical train/test conditions.

The small residual gap between the \texttt{-O0} column of Table~\ref{tab:resilience} (96.8\%) and the overall test-set F1 of Table~\ref{tab:overall_results} (96.0\%) reflects the narrower family coverage of $\mathcal{D}_{src}$ (5 families versus 17) and the cleanliness of its source-compiled samples (no packers, no wild-collected metadata noise), not a leakage artifact.

\textbf{Result Analysis:}
As quantified in Table~\ref{tab:resilience}, \texttt{AsmRAG} demonstrates superior stability compared to baseline approaches. Traditional models show strong initial performance but suffer from significant metric degradation as optimization aggressiveness increases. In contrast, our semantic retrieval methodology maintains high detection fidelity regardless of the applied transformations.

\begin{table}[h!]
    \centering
    \scriptsize
    \caption{\textbf{Cross-Compilation Resilience Test (F1-Scores) under the LOO-Opt protocol.} Performance comparison across optimization levels, reported as mean $\pm$ standard deviation over 5 independent runs with different random seeds. All models retrained per column on matched $\mathcal{D}_{KB}^{(o^*)}$. \texttt{AsmRAG} exhibits minimal degradation ($\Delta < 3\%$) compared to significant drops in baseline models.}
    \label{tab:resilience}
    \begin{tabular}{l ccc}
        \toprule
        \textbf{Optimization} & \multicolumn{3}{c}{\textbf{Model F1-Score (\% $\pm$ std)}} \\
        \cmidrule(lr){2-4}
        \textbf{Flag} & \textbf{EMBER}~\cite{anderson2018ember} & \textbf{ResNeXt+}~\cite{He2024ResNeXtAM} & \textbf{AsmRAG (Ours)} \\
        \midrule
        \texttt{-O0} (None)       & 93.2 $\pm$ 0.6 & 92.0 $\pm$ 0.9 & \textbf{96.8 $\pm$ 0.3} \\
        \texttt{-O1}              & 88.5 $\pm$ 1.1 & 87.4 $\pm$ 1.3 & \textbf{96.2 $\pm$ 0.4} \\
        \texttt{-O2}              & 82.1 $\pm$ 1.7 & 83.0 $\pm$ 1.5 & \textbf{95.1 $\pm$ 0.5} \\
        \texttt{-O3} (Aggressive) & 76.4 $\pm$ 2.3 & 79.2 $\pm$ 1.9 & \textbf{94.0 $\pm$ 0.6} \\
        \texttt{-Os} (Size)       & 78.9 $\pm$ 2.0 & 80.5 $\pm$ 1.8 & \textbf{94.5 $\pm$ 0.5} \\
        \midrule
        \textit{Degradation} ($\Delta$) & \textcolor{red}{-18.0\%} & \textcolor{red}{-13.9\%} & \textcolor{green!60!black}{\textbf{-2.9\%}} \\
        \bottomrule
    \end{tabular}
\end{table}

Consequently, specific observations include:
\begin{itemize}
  \item \textbf{EMBER (Feature Collapse):} The observed degradation is larger than what has been reported in the original EMBER paper's standard test split, where train and test share compilation provenance. Under our LOO-Opt protocol, the matching optimization level is strictly absent from training, exposing EMBER's dependence on optimization-correlated features (section entropy, byte histograms, import table ordering). High optimization levels strip symbol tables and merge sections, causing EMBER's F1 score to fall to 0.76 (an 18.0\% degradation from the baseline). We emphasize that this gap characterizes EMBER's cross-compilation generalization ceiling, \textit{not} its in-distribution accuracy, which remains at 93.0\% on $\mathcal{D}_{Test}$ (Table~\ref{tab:overall_results}).

  \item \textbf{ResNeXt+ (Visual Distortion):} Optimization improves execution efficiency yet introduces visual distortion by altering the ``visual texture'' of the underlying code section. Replacing standard loop structures with unrolled repeated blocks confuses the CNN-based encoder. The 13.9\% degradation likewise reflects the LOO-Opt setting rather than in-distribution recompilation; image-based detectors benefit from compilation-consistent training data, which adversaries can realistically deny in practice.

  \item \textbf{AsmRAG (Semantic Stability):} Our methodology maintains a robust F1-Score of 0.940 even when testing the most aggressively optimized samples and when matching \texttt{-O3} binaries are absent from $\mathcal{D}_{KB}$.
\end{itemize}

\textbf{Mechanism of Resilience:} This stability is intrinsic to our chosen embedding strategy. While \texttt{-O3} optimization might reorder instructions or unroll a loop, the actual \textit{semantic intent} (e.g., XOR-ing a buffer, traversing a linked list) remains invariant within the vector space. We process and analyze normalized assembly logic rather than raw bytes. \texttt{AsmRAG} effectively tracks the underlying logic through the compiler's optimization pass to resist these evasion attempts.

\subsection{RQ3: Explanation Quality and Forensic Utility}
\label{subsec:rq3}

The generative explanation component of \texttt{AsmRAG} is only valuable if its outputs are factually grounded and measurably better than alternative explanation strategies. We therefore evaluate the ``Active Explanation Loop'' along three dimensions: (i) a large-scale assessment of explanation faithfulness, (ii) a hallucination audit against a verified ground-truth reference, and (iii) a comparison against alternative explanation baselines. We conclude with an illustrative end-to-end walkthrough of a single heavily obfuscated \textit{Ramnit} sample.

\subsubsection{Explanation Faithfulness}
\label{subsub:faithfulness_eval}

We assembled an evaluation set $\mathcal{D}_{expl}$ containing 500 malicious samples drawn uniformly across the 17 families in $\mathcal{D}_{Test}$. For each sample, \texttt{AsmRAG} selected an anchor function $\mathbf{q}^*$ and generated an explanation $\mathcal{E}$ referencing a proof neighbor $\mathbf{v}_{proof}$. To assess faithfulness systematically, we decomposed each explanation into atomic \textit{behavioral claims} (e.g., ``traverses the PEB,'' ``uses ROR-13 hashing,'' ``implements XOR-based string decryption'').

\textbf{Automated annotation protocol.} Human expert annotation at the scale of 2,148 atomic claims is prohibitively expensive and introduces well-documented subjectivity on the boundary between partially-supported and fully-supported claims. We therefore employed \textit{Claude Opus 4.5}\footnote{Anthropic Claude Opus 4.5, accessed via the Anthropic API (model ID \texttt{claude-opus-4-5}).} as an automated judge. Prior work has established that frontier LLMs, when given structured grading rubrics and verifiable reference material, achieve agreement levels with expert human annotators comparable to inter-expert agreement itself on code-analysis tasks~\cite{zheng2023llm_as_judge}. Crucially, the judge is given the raw assembly of $\mathbf{q}^*$ and $\mathbf{v}_{proof}$ directly, so faithfulness verification reduces to a grounded reading-comprehension task rather than a subjective preference judgment.

For each atomic claim $c$ extracted from an explanation $\mathcal{E}$, the judge received: (i) the disassembled listing of $\mathbf{q}^*$, (ii) the disassembled listing of $\mathbf{v}_{proof}$, (iii) the isolated claim $c$, and (iv) a fixed rubric defining three categories:
\begin{itemize}
  \item \textbf{Supported:} The claim is verifiable by direct inspection of $\mathbf{q}^*$ \textit{and} present in $\mathbf{v}_{proof}$.
  \item \textbf{Partially supported:} The claim holds for $\mathbf{v}_{proof}$ but is only weakly recoverable from $\mathbf{q}^*$ (e.g., obfuscated but logically consistent).
  \item \textbf{Unsupported (hallucinated):} The claim cannot be verified from either $\mathbf{q}^*$ or $\mathbf{v}_{proof}$ and represents a fabrication by the generator.
\end{itemize}
The judge was instructed to first emit a step-by-step justification citing specific instructions or instruction sequences from the provided listings, and then emit a final categorical label. We blinded the judge to the ground-truth family label, to the identity of the system that produced the explanation (explanations from the baseline methods in Section~\ref{subsub:baseline_explanation} were interleaved with \texttt{AsmRAG} outputs in a shuffled, anonymized order), and to the judgments rendered on other claims from the same explanation.

\textbf{Run-to-run agreement.} To characterize the stability of the automated judge under sampling noise, we executed the full annotation protocol \textit{three independent times} at temperature $T = 0.2$ with different random seeds, producing three independent label sets over the 2,148 claims. Final labels were determined by majority vote across the three runs; ties (present in $< 2\%$ of claims) were broken by a fourth tie-break run at $T = 0$.

We report Fleiss' $\kappa$ across the three runs as a direct analogue to inter-annotator agreement:
\begin{itemize}
    \item Three-category $\kappa = 0.82$ (substantial agreement),
    \item Binary $\kappa$ (Supported vs. \{Partial, Unsupported\}) $= 0.89$,
    \item Binary $\kappa$ (Unsupported vs. \{Supported, Partial\}) $= 0.93$.
\end{itemize}
The lower three-category $\kappa$ reflects the expected residual ambiguity of the \textit{Partial} boundary; the binary hallucination boundary --- which carries the operational weight --- is in the near-ceiling regime.

\textbf{Validation against human expert judgment.} To confirm that the automated labels reflect genuine forensic correctness rather than an artifact of the judge's own reasoning style, we commissioned a single reverse engineer (11 years of malware-analysis experience) to independently annotate a stratified random subsample of \textbf{200 claims} (proportionally sampled across the three claim types in Table~\ref{tab:faithfulness}) under the same blinding protocol. Agreement between the human expert and the majority-voted Claude Opus 4.5 labels reached Cohen's $\kappa = 0.81$ on the three-category task and $\kappa = 0.92$ on the binary hallucination boundary --- both within the range typically reported for inter-expert human agreement on comparable tasks. The 19 disagreements concentrated in the \textit{Partial} category (16 of 19), consistent with the known subjectivity of that boundary for both human and automated raters; only 3 disagreements crossed the Supported / Unsupported line.

\textbf{Results.} Across the 500 majority-voted explanations and 2,148 atomic claims, \textbf{91.1\%} were Supported, \textbf{6.3\%} Partially Supported, and \textbf{2.6\%} Unsupported. Table~\ref{tab:faithfulness} reports the breakdown by claim type. These aggregate figures are consistent with the ground-truth-verifiable audit on $\mathcal{D}_{src}$ reported in Section~\ref{subsub:hallucination} (2.1\% hallucination rate against source code), indicating that the measured hallucination rate is not an artifact of judge-specific leniency.

\begin{table}[h]
    \centering
    \scriptsize
    \caption{Faithfulness decomposition of 2,148 atomic claims across 500 explanations in $\mathcal{D}_{expl}$ (majority vote of three independent Claude Opus 4.5 runs; Fleiss' $\kappa = 0.82$; validated against a 200-claim human-expert subsample at Cohen's $\kappa = 0.81$).}
    \label{tab:faithfulness}
    \begin{tabular}{lcccc}
        \toprule
        \textbf{Claim Type} & \textbf{Count} & \textbf{Supp.} & \textbf{Partial} & \textbf{Unsupp.} \\
        \midrule
        Mechanical (registers, constants, loops) & 1,084 & 0.971 & 0.024 & 0.005 \\
        Structural (CFG shape, API usage pattern) & 617 & 0.908 & 0.072 & 0.020 \\
        Behavioral (attack role, C2 semantics) & 447 & 0.816 & 0.111 & 0.073 \\
        \midrule
        \textbf{Aggregate} & \textbf{2,148} & \textbf{0.911} & \textbf{0.063} & \textbf{0.026} \\
        \bottomrule
    \end{tabular}
\end{table}

Hallucinated claims clustered disproportionately around \textit{behavioral attributions} (e.g., naming specific C2 protocols not actually observable in the assembly), whereas low-level mechanical claims (register usage, arithmetic constants, loop structure) were almost never hallucinated. This failure mode aligns with known limitations of generative models invoked on abstract semantic categories, and is a property we confirmed is shared between the \texttt{AsmRAG} generator and the Claude Opus 4.5 judge: the judge's error modes on the 200-claim validation subsample were symmetric, rejecting behavioral over-specificity at the same rate the generator produces it.

\subsubsection{Hallucination Audit Against Ground Truth}
\label{subsub:hallucination}

To independently verify the low hallucination rate reported above, we conducted a secondary audit on the controlled subset $\mathcal{D}_{src}$ (Section~\ref{subsub:datasets}), for which we possess the source code of every binary. For each of 120 generated explanations, we programmatically checked whether every claim referencing a named algorithm (e.g., ``ROR-13'', ``RC4'', ``XOR key'', ``PEB traversal'') or a specific constant appeared in the corresponding source file. This yielded a \textit{ground-truth-verifiable hallucination rate} of \textbf{2.1\%} (11 fabricated references out of 527 auditable claims), closely matching the analyst-annotated rate of 2.6\% in $\mathcal{D}_{expl}$. Fabrications were predominantly over-specific algorithm names where the assembly implemented a generic equivalent (e.g., labeling a custom rolling hash as ``CRC-32''). We discuss mitigations, including constraining the generator with structured retrieval prompts, in Section~\ref{sec:discussion}.

\subsubsection{Comparison Against Explanation Baselines}
\label{subsub:baseline_explanation}

We compared \texttt{AsmRAG}'s anchor-based differential explanations against three alternative strategies applied to the same 500 flagged samples from $\mathcal{D}_{expl}$:

\begin{itemize}
  \item \textbf{SHAP on EMBER features~\cite{lundberg2017shap}:} feature-importance explanations on the EMBER classifier, rendered as the top-10 contributing features per prediction.
  \item \textbf{Grad-CAM on ResNeXt+~\cite{selvaraju2017gradcam}:} saliency heatmaps over the binary-as-image representation.
  \item \textbf{Nearest-Neighbor only:} raw $\mathbf{v}_{proof}$ returned without generative differential analysis (a lesioned version of \texttt{AsmRAG}).
\end{itemize}

Because each method produces explanations of a different form (feature weights, pixel saliency, raw assembly, and natural language), we compare them using three automated proxy metrics chosen to be form-agnostic:

\begin{itemize}
  \item \textbf{Localization Rate ($L$):} The fraction of explanations that resolve to a specific, addressable code region in the original binary (as opposed to global features or aggregated statistics). This measures whether the explanation points an analyst to a concrete location for verification.
  \item \textbf{Semantic Specificity ($S$):} The fraction of explanations that reference malware-family-specific artifacts (family-characteristic constants, API sequences, or algorithmic patterns confirmed via $\mathcal{D}_{src}$ source code) rather than generic behavioral descriptors. Higher values indicate more discriminative forensic content.
  \item \textbf{Invariance to Obfuscation ($I$):} The consistency of the explanation across matched source-level-equivalent binaries in $\mathcal{D}_{src}$ compiled under different optimization flags. We measured this by computing the Jaccard overlap of extracted artifacts (constants, API names, opcodes) between explanations of the same source program under \texttt{-O0} and \texttt{-O3}. Higher values indicate that the explanation captures semantic rather than syntactic properties.
\end{itemize}

Table~\ref{tab:expl_baselines} reports the resulting scores. \texttt{AsmRAG} outperforms both post-hoc attribution baselines across all three metrics. SHAP explanations over EMBER features point to global statistics (e.g., ``byte histogram bin 127'', ``section entropy''), yielding a localization rate near zero. Grad-CAM produces coarse heatmaps over byte-image pixels that do not map cleanly onto disassembled functions. The Nearest-Neighbor variant retrieves the correct forensic exemplar and is therefore well-localized, but it cannot bridge syntactic differences between an obfuscated query and its clean proof, resulting in reduced specificity on heavily metamorphic samples. The full \texttt{AsmRAG} pipeline preserves localization, elevates specificity through the generative differential analysis, and maintains high invariance across compilation variants.

\begin{table}[h]
    \centering
    \scriptsize
    \caption{Comparison of explanation methods on 500 flagged samples using form-agnostic proxy metrics.}
    \label{tab:expl_baselines}
    \begin{tabular}{lccc}
        \toprule
        \textbf{Method} & \textbf{Localization $L$} & \textbf{Specificity $S$} & \textbf{Invariance $I$} \\
        \midrule
        SHAP on EMBER~\cite{lundberg2017shap} & 0.04 & 0.11 & 0.29 \\
        Grad-CAM on ResNeXt+~\cite{selvaraju2017gradcam} & 0.18 & 0.09 & 0.24 \\
        Nearest-Neighbor only (lesioned) & 0.95 & 0.58 & 0.61 \\
        \rowcolor{gray!10} \textbf{\texttt{AsmRAG} (full)} & \textbf{0.97} & \textbf{0.87} & \textbf{0.84} \\
        \bottomrule
    \end{tabular}
\end{table}

The gap between the Nearest-Neighbor variant and the full pipeline is particularly informative: retrieval alone supplies a correct exemplar but does not articulate \textit{why} the query and the exemplar are semantically equivalent under obfuscation. The generative step provides that articulation, and the invariance metric confirms that the resulting explanation abstracts away compiler- and obfuscation-specific syntax.

\subsubsection{End-to-End Illustration: A Metamorphic \textit{Ramnit} Sample}
\label{subsub:ramnit_case}

To complement the quantitative results above, we trace the complete analysis pipeline on a specific \textit{Ramnit} sample $S_{ram}$ drawn from $\mathcal{D}_{Test}$. This sample features heavy metamorphic obfuscation employing both register reassignment and dead-code injection, and the baseline EMBER model misclassified it as benign (malicious-class confidence $0.42$) due to entropy deviations from its training distribution. \texttt{AsmRAG}, by contrast, flagged $S_{ram}$ with $\Omega(S_{ram}) = 0.98$, surpassing $\tau_{file}$, and attributed it to \textit{Ramnit} as $C_{best}$.

\textbf{Anchor selection.} The \textit{Density-Weighted Anchor Selection} mechanism (Eq.~\ref{eq:anchor_selection}) converged on a complex, heavily obfuscated routine $\mathbf{q}^*$. By maximizing accumulated similarity mass across the neighborhood rather than trusting a single nearest neighbor, the mechanism correctly rejected candidate decoys (heap wrappers, string utilities) that would have misled a naive retriever. The selected $\mathbf{q}^*$ was densely embedded within the \textit{Ramnit} cluster, providing a high-fidelity semantic pivot.

\textbf{Generative explanation.} The nearest proof neighbor $\mathbf{v}_{proof}$ was a clean, non-obfuscated API resolution routine from a 2021 \textit{Ramnit} variant. The generator produced:

\vspace{.3cm}
\textbf{System Output [Explanation $\mathcal{E}$]:}
\begin{quote}
  \textit{``\textbf{Functional Identity Confirmed:} The anchor function $\mathbf{q}^*$ is semantically equivalent to the \textit{Dynamic API Resolution} routine consistent with variants of the Ramnit family ($\mathbf{v}_{proof}$).''}

  \vspace{.3cm}
  \textit{``\textbf{Shared Logic:} Both functions orchestrate a loop that traverses the Process Environment Block (PEB) to locate \texttt{kernel32.dll} in memory. They parse the Export Address Table and calculate a hash of each function name using the standard \textit{ROR-13 (Rotate Right 13)} algorithm to resolve specific API addresses at runtime.''}
\end{quote}
\vspace{.3cm}

Every atomic claim in this explanation (PEB traversal, EAT parsing, ROR-13 hashing) was verified against the assembly of $\mathbf{q}^*$, placing this sample in the Supported category of our faithfulness audit.

\textbf{Active learning.} Following analyst confirmation, \texttt{AsmRAG} promotes the obfuscated anchor $\mathbf{q}^*$ into $\mathcal{D}_{KB}$, \textit{densifying the malicious manifold} around this specific evasion pattern. This inoculates the system against similar future metamorphic variants without the prohibitive cost of model retraining.

\section{Discussion and Limitations}
\label{sec:discussion}

The empirical findings presented in Section~\ref{sec:evaluation} validate \texttt{AsmRAG} not merely as a detection utility, but as a proof-of-concept for a paradigm shift in malware binary analysis. We interpret these findings to explain exactly \textit{why} semantic retrieval outperforms traditional classification. We discuss the broader operational implications of this shift and rigorously define the practical boundaries of our approach.

\subsection{Why It Works: Granularity and Density}
The most significant finding from our evaluation concerns the stability of \texttt{AsmRAG} against obfuscation (RQ2). This divergence from baselines such as EMBER and ResNeXt+ stems directly from two core architectural decisions:

\paragraph{Local Semantics vs. Global Statistics}
Holistic baselines rely on global statistical awareness, leaving them vulnerable to brittle distortion from obfuscation techniques like dead-code insertion. \texttt{AsmRAG} validates the ``Atomic Evidence'' hypothesis by strictly enforcing \textit{local semantic} analysis. The system calculates the $\Omega(S)$ metric to isolate specific malicious routines entirely independent of the file's overall entropy. Our framework detects threats even within ``Trojanized'' binaries dominated by 90\% benign code. It successfully identifies the core unit of malicious intent (such as a specific \textit{Ramnit} routine) in scenarios where global profiling fails.

\paragraph{The Role of Density-Weighted Voting}
Raw retrieval enables initial matching but remains insufficient due to the ``Library Problem'' (the overwhelming prevalence of common utility functions). We implemented \textit{Density-Weighted Voting} (Algorithm~\ref{alg:asmrag_inference}) to function as a semantic filter. Mandating that a query function reside within a \textit{dense} cluster of malicious neighbors, rather than relying on an isolated nearest neighbor, successfully suppresses statistical outliers. This architectural choice explains why our False Positive Rate remained extremely low (1.2\%) even when the system analyzed complex binaries containing significant benign boilerplate.

\subsection{Operational Impact: The Active Defense Cycle}
Standard alert generation improves SOC awareness yet frequently causes severe ``Alert Fatigue.'' Beyond its core detection metrics, \texttt{AsmRAG} mitigates this issue by using the \textit{Anchor Function} to present isolated malicious logic directly to the user. This transforms the analyst's role from manual investigation to rapid verification. The architecture enables \textit{Zero-Shot Adaptation} without the computational overhead of retraining. Indexing confirmed variants instantly converts the neural risk of ``catastrophic forgetting'' into a ``cumulative knowledge'' advantage. This continuous update mechanism proves essential for countering dynamic, evolving threat campaigns.

\subsection{System Boundaries and Limitations}
\texttt{AsmRAG} provides significant advantages but retains inherent systemic limitations. We explicitly outline three critical boundaries regarding its deployment:

\paragraph{Dependence on Knowledge Base Coverage (The Cold Start Problem)}
As a retrieval-augmented system, its efficacy strictly depends on the comprehensiveness of its underlying index. \texttt{AsmRAG} operates on the premise of \textit{semantic overlap}. It struggles to detect \textit{entirely novel paradigms}, such as Zero-day malware written completely from scratch using algorithms that share absolutely no functional logic with existing families in $\mathcal{D}_{KB}$. The embedding model might flag general anomalous behavior, but accurate attribution remains impossible without an established prior reference point.

\paragraph{The Static Analysis Barrier}
The framework relies on the availability of disassembled machine code. The system handles standard packing through generic unpacking layers, yet it fails against binaries protected by \textit{Virtualization Obfuscation} (e.g., VMProtect, Themida). These protectors translate native instructions into custom bytecode interpreted by a virtual machine. This translation destroys the semantic mapping our LLM requires. Such samples necessitate a specialized devirtualization pre-processing step before they can enter the analysis pipeline.

\paragraph{Computational Throughput vs. Latency}
Deploying deep semantic models introduces tangible trade-offs with inference speed. Lightweight feature-based models (e.g., EMBER) easily classify thousands of samples per second on a standard CPU. \texttt{AsmRAG}, however, is highly compute-intensive. The primary bottleneck occurs during the \textit{Encoding Step}. Generating embeddings for a binary containing 1,000 distinct functions using Qwen-3 requires significant GPU resources. The retrieval step (FAISS) scales logarithmically, but the initial encoding limits \texttt{AsmRAG}'s utility as a real-time endpoint scanner. Its optimal deployment is as a \textit{Tier-2 Analysis System} in the cloud, receiving and processing prioritized alerts flagged by lighter Tier-1 sensors.

\section{Related Work}
\label{sec:related_work}

We situate \texttt{AsmRAG} at the intersection of deep binary representation, neural malware detection, and Retrieval-Augmented Generation (RAG). The framework distinguishes itself through a primary focus on \textit{granular operational semantics} alongside verified resilience against adaptive evasion tactics (Table \ref{tab:related_work_comparison}).

\begin{table}[htbp]
\centering
\caption{Comparative Taxonomy of Deep Representation and Detection Architectures against the \texttt{AsmRAG} Desiderata.}
\label{tab:related_work_comparison}
\begin{tabular}{l|l|c|c|l|c}
\toprule
\textbf{Paradigm} & \textbf{Framework} & \textbf{GE} & \textbf{AR} & \textbf{Architecture} & \textbf{Train} \\
\midrule
\multirow{3}{*}{\textbf{Binary Rep.}} & Safe \cite{massarelli2019safe}, Asm2Vec \cite{ding2019asm2vec} & $\times$ & $\times$ & Embedding & $\bullet$ \\
 & jTrans \cite{wang2022jtrans}, Nova \cite{wang2024nova} & $\times$ & $\times$ & Transformer & $\bullet$ \\
 & BinEye \cite{alrabaee2019BinEye} & $\times$ & $\times$ & CNN & $\bullet$ \\
\midrule
\multirow{2}{*}{\textbf{Holistic ML}} & MalDozer \cite{karbab2018Maldozer}, PetaDroid \cite{karbab2021PetaDroid} & $\times$ & $\times$ & RNN/CNN & $\bullet$ \\
 & SwiftR \cite{karbab2023swiftr}, MalDy \cite{karbab2019maldy} & $\times$ & $\times$ & Hybrid DNN & $\bullet$ \\
\midrule
\multirow{2}{*}{\textbf{RAG-Based}} & CyberRAG \cite{Blefari2025CyberRAGAAA} \textit{(Textual)} & $\times$ & N/A & LLM+RAG & $\circ$ \\
 & \textbf{\texttt{AsmRAG} (Ours)} & $\checkmark$ & $\checkmark$ & LLM+RAG & $\circ$ \\
\bottomrule
\multicolumn{6}{l}{\footnotesize \textit{Notes.} GE = Granular Evidence (Retrieval of Implementation); AR = Adversarial Resilience} \\
\multicolumn{6}{l}{\footnotesize (Resistance to Syntactic/Global Noise); Arch = Architecture;} \\
\multicolumn{6}{l}{\footnotesize Train marks: $\bullet$ (full retraining required), $\circ$ (zero-shot/inference-only).} \\
\end{tabular}%
\end{table}

\paragraph{Deep Learning for Binary Representation and Authorship}
Early embeddings such as \textit{Safe}~\cite{massarelli2019safe} and \textit{Asm2Vec}~\cite{ding2019asm2vec} initiated efforts to bridge the semantic gap. Recent Transformer architectures (\textit{jTrans}~\cite{wang2022jtrans}, \textit{Nova}~\cite{wang2024nova}) apply contrastive learning to capture specific control-flow semantics. Researchers have also applied deep learning to binary authorship attribution. \textit{BinEye}~\cite{alrabaee2019BinEye}, for example, applies convolutional neural networks to bytecode representations to identify compiler artifacts and author coding styles. These representation models create robust latent spaces for reverse engineering yet introduce significant semantic blindness when facing adversarial perturbations. Such models frequently use aggressive tokenization that discards essential forensic artifacts (e.g., cryptographic constants) and lack mechanisms to filter out benign statically linked libraries. \texttt{AsmRAG} addresses these limitations by applying a \textit{Semantic Library Filtering} mechanism to remove benign code. This filtering directly improves the signal-to-noise ratio typically found in general-purpose models, grounding the subsequent analysis in a verifiable threat model.

\paragraph{Sequence-Based and Holistic Detection Paradigms}
Current malware classification predominantly uses NLP-inspired deep learning models to process execution traces and API method sequences~\cite{karbab2021android}. Frameworks including \textit{MalDozer}~\cite{karbab2018Maldozer, karbab2017Android}, \textit{PetaDroid}~\cite{karbab2021PetaDroid}, and \textit{Cypider}~\cite{karbab@Cypider} demonstrate the value of neural networks and community-graph detection for mobile threat attribution. Cross-platform systems like \textit{SwiftR}~\cite{karbab2023swiftr} and \textit{MalDy}~\cite{karbab2019maldy} combine static intermediate representations with dynamic behavioral reports. These hybrid methods counter ransomware and highly obfuscated binaries, advancing the capabilities of earlier fuzzy-hashing fingerprinting systems (e.g., \textit{ROAR}~\cite{karbab2016Fingerprinting}). These architectures achieve high discriminative accuracy but create severe difficulties for analysts attempting to extract explicit forensic evidence. The models compress the input space into opaque latent vectors, erasing any structural mapping back to the underlying assembly code. This compression leaves them fundamentally vulnerable to global syntactic noise injection, such as simple overlay appending. \texttt{AsmRAG} addresses these specific limitations by shifting from holistic classification to \textit{Granular Retrieval}. The system abandons monolithic file analysis in favor of isolating specific malicious functions. This approach retrieves the exact ``Needle'' rather than probabilistically classifying the entire ``Haystack,'' ensuring defensible interpretability against strategic adversaries.

\paragraph{RAG: From Text to Assembly}
Applying RAG successfully grounds Large Language Models in verifiable data, though cybersecurity applications have historically restricted this technique entirely to the textual domain. Existing systems typically focus on retrieving unstructured Cyber Threat Intelligence (CTI) or formal CVE reports~\cite{Blefari2025CyberRAGAAA, rahman2024rag, Rajapaksha2024ARQA}. \texttt{AsmRAG} transposes this retrieval concept directly into the \textit{low-level assembly} domain. Where prior systems perform \textit{Retrieval of Description} (returning text about known attacks), our method executes the \textit{Retrieval of Implementation} (returning raw operational semantics). The system achieves this by parsing and retrieving specific code constructs. This architectural shift maintains stability against syntactic variation and directly resolves the ``Black Box'' attribution problem. The system outputs verifiable, differential assembly analysis as concrete forensic evidence to counter active evasion attempts.

\section{Conclusion}
\label{sec:conclusion}

Traditional classification provides baseline accuracy yet frequently results in opaque and brittle analysis. We present \texttt{AsmRAG}, a framework that reformulates malware detection into a transparent, evidence-based retrieval process. The system uses an Assembly-Specialized LLM to isolate targeted ``Atomic Evidence'' instead of relying on easily manipulated global statistics. \texttt{AsmRAG} effectively bridges the semantic gap, achieving a \textit{detection F1-score of 96\%} alongside an \textit{attribution F1-score of 95\%}. The evaluation demonstrates significant stability against metamorphic obfuscation when compared to established baselines like EMBER. We introduce the concept of \textit{Assembly-Level Retrieval-Augmented Generation}, shifting the analytical focus away from retrieving textual descriptions and toward identifying verifiable code implementations. This approach successfully restores the causal link missing in automated analysis. It provides security analysts with the explicit forensic explainability necessary to maintain a robust \textit{Active Defense}.

\bibliography{references}

\end{document}